\documentclass[aps,pra,twocolumn,amsfonts,amsmath,amssymb,floatfix,showpacs,a4paper,10pt]{revtex4-1}
\usepackage{bm}
\usepackage{epsfig}
\usepackage[english]{babel}
\usepackage{graphicx}
\usepackage{longtable}
\def \hmol{H$_2$}%
\def \htpl{H$_{2}^{+}$}%
\def \bsig {B\,$^1\Sigma_{u}^{+}$}%
\def \xsig {X\,$^1\Sigma_{g}^{+}$}%
\def \iong {H$_2$$^+$\, X\,$^2\Sigma_{g}$}%
\def \schr{Schr\"odinger}
\def \wcm{W/cm$^2$}
\def \op#1{\hat{\rm #1}}
\def \susu{$|1\sigma_u \rangle |1\sigma_u \rangle$}
\begin{document}

%
%
%

\title{Break-down of the single-active-electron approximation for 
       one-photon ionization of the B\,$\bm{^1\Sigma_u^+}$ state of 
       H$_{\bm 2}$ exposed to intense laser fields} 

\author{Manohar Awasthi}
\author{Alejandro Saenz}
\affiliation{Humboldt-Universit\"at zu Berlin, Institut f\"ur Physik,
AG Moderne Optik, Newtonstr.~15, 12489 Berlin, Germany.}

\email{Alejandro.Saenz@physik.hu-berlin.de}

\date{\today}

\begin{abstract} 
  Ionization, excitation, and de-excitation to the ground state is studied
  theoretically for the first excited singlet state {\bsig} of {\hmol} exposed
  to intense laser fields with photon energies in between about 3\,eV and
  13\,eV. A parallel orientation of a linear polarized laser and the molecular 
  axis is considered.  Within the dipole and the fixed-nuclei
  approximations the time-dependent Schr\"odinger equation describing the
  electronic motion is solved in full dimensionality and compared to simpler 
  models. A dramatic break-down 
  of the single-active-electron approximation is found and explained to be 
  due to the inadequate description of the final continuum states. 
\end{abstract}


\maketitle


\section{Introduction}\label{sec:intro} 

Recently, there has been an increased interest in the development of intense
light sources with wavelengths shorter than the one of the titanium:sapphire
lasers. This includes both high-harmonic sources and, especially, the
free-electron lasers (FEL)~\cite{gen:made71}.  Although the primary goal is to
achieve radiation in the x-ray regime, also the vacuum UV is of some practical
interest.  Interesting results have recently been achieved, e.\,g., at the
free-electron laser FLASH in Hamburg~\cite{sfa:rude08, sfa:rich09, sfm:jian09,
  sfa:zhu09}.

For a laser pulse with a photon frequency centered at $\omega$ and a peak
electric field $F_0$ the Keldysh parameter $\gamma =
\sqrt{I_p/U_p}$~\cite{sfa:keld65} with the ionization potential $I_p$ and the
ponderomotive energy $U_p = \frac{{F_0}^2}{4 \omega^2}$ is usually used as a
(rough) criterion to distinguish the so-called multiphoton ($\gamma>1$) and
tunneling ($\gamma<1$) regimes. With the large photon energies as they are
available from FELs one is likely to remain in the multiphoton
regime. Interesting questions in this context are whether at these photon
energies even lowest-order perturbation theory (LOPT) may be sufficient to
properly describe most ionization processes in such laser pulses and at which
intensities non-perturbative behavior sets in. A natural candidate for
investigating these questions is H$_2$, since its two electrons exposed to
intense laser pulses can be treated perturbatively
\cite{sfm:apal01,sfm:apal02} and non-perturbatively
\cite{sfm:kawa00,sfm:awas05,sfm:awas06,sfm:pala06,sfm:vann08,sfm:vann09}
within full dimensionality.

Despite the fact that electronically excited states have been found
theoretically and experimentally to be populated by a non-negligible amount in
intense laser pulses \cite{sfa:debo92,sfa:jone93, sfa:hert97, sfa:mull99,
  sfa:nubb08, sfm:awas08, sfm:mans09}, most theoretical studies on H$_2$
exposed to laser fields have concentrated on the ground state as initial
state. However, the field-induced coupling
to the typically much more closely spaced neighbor states may 
complicate the strong-field behavior of excited states. On the other hand, 
since higher photon fluxes require lower target densities, the novel 
intense light sources may allow for direct studies of electronically 
excited states similar as for ions. Such an investigation was, e.\,g., 
recently demonstrated for HeH$^+$ \cite{dia:pede07,dia:dumi09}.

To the authors' knowledge, the only theoretical studies of excited states of
H$_2$ in intense laser fields were reported in \cite{sfm:saen02c} in which
their behavior in a (quasi)static field was investigated and in 
\cite{sfm:bart08} in which one-photon ionization of the {\bsig} state was 
considered within a newly developed single-active-electron (SAE) approximation 
based on Koopmans' picture.  In fact, the latter work has motivated the
present investigation.  Thus the
first electronically excited state {\bsig} of {\hmol} exposed to intense laser
fields is considered and the validity of the SAE approximation for this
excited state is investigated. The {\hmol} molecule is treated in the
fixed-nuclei approximation at $R\,=\,2.2998\,a_0\,$ with the field
polarization being parallel to the internuclear axis.  First, 8\,eV photons
were considered as in \cite{sfm:bart08}. Motivated by the found order of 
magnitude deviation between the present full two-electron calculation 
and the one in \cite{sfm:bart08}, a larger photon-energy range
within the regime of one-photon ionization is considered, but only for the
case of a single open channel, i.\,e., before the photon energy is 
sufficient to leave the H$_2^+$ ion in an electronically excited state. 
It may be noted that some of the photon energies used in the calculations 
are already available with FELs~\cite{gen:andr00, gen:yu03}.

After a brief description of the methods in Sec.\,\ref{sec:method} the results
are presented and discussed in Sec.\,\ref{sec:results}. This includes a
discussion of the results for a photon energy of 8\,eV in Sec.\,\ref{ss:om8ev} 
and for variable photon energies in Sec.\,\ref{ss:sinphoion}. A simple model 
for explaining the failure of the SAE is given in Sec.\,\ref{ss:explain}, 
followed by a conclusion in Sec.\,\ref{sec:conclusion}. 
Atomic units are used, if not stated otherwise.      
%
%
%
%
\section{Method}\label{sec:method}
Most results of this work have been obtained by a full-dimensional  
solution of the time-dependent Schr\"odinger equation (TDSE) describing 
the two electrons of H$_2$ exposed to an intense laser pulse with 
linear polarization. The molecular axis is assumed to be aligned 
parallel to the electric field of the laser. The semi-classical 
non-relativistic dipole approximation is adopted in which the 
laser field is described classically, while the molecular system is 
treated quantum mechanically. The shown results are obtained for 
fixed internuclear separations. The details of the approach have 
been given previously \cite{sfm:awas05,sfm:awas06,sfm:awas08,sfm:vann08} 
and are thus only briefly repeated.  

The electronic TDSE describing {\hmol} in a laser field is given within 
the above-mentioned approximations as
\begin{equation}\label{eq:fulltdse}
   \imath \frac{\partial}{\partial t} \Psi({\bf r}_1,{\bf r}_2,t) 
           \, = \, \left( \op{H}_0 \, + \, \op{D} (t) \right) 
                                    \Psi({\bf r}_1,{\bf r}_2,t) \quad 
\end{equation}
where $\op{H}_0$ is the field-free electronic Hamiltonian of H$_2$ 
and $\op{D} (t)$ describes the
interaction with the field. This interaction may be given in either 
velocity form, $\op{D}(t) = {\bf A}(t)\cdot({\bf p}_1 + {\bf p}_2)$,  
or in length form,  $\op{D}(t) = -{\bf F}(t)\cdot({\bf r}_1+{\bf r}_2)$,  
where ${\bf p}_i$ is the momentum operator of electron $i$ while 
${\bf A}$ and ${\bf  F}$ represent vector 
potential and the electric field, respectively. 

The wavefunction  
\begin{equation}\label{eq:expansionpsi}
     \Psi ({\bf r}_1,{\bf r}_2, t) \:=\: 
             \sum_{nL} \: b_{nL}(t)\, \phi_{nL}({\bf r}_1,{\bf r}_2)
\end{equation}
is expanded in terms of the field-free states $ \phi_{nL}$ and the 
time-dependent coefficients $b_{nL}(t)$. The field-free wavefunctions 
$\phi_{nL}({\bf r}_1,{\bf r}_2)$ and corresponding energy eigenvalues 
$E_{nL}$ are obtained from the solution of the time-independent  
{\schr} equation (TISE),
\begin{equation}\label{eq:h2tise}
  \op{H}_0  \phi_{nL}({\bf r}_1,{\bf r}_2) \:=\: 
                E_{nL} \phi_{nL}({\bf r}_1,{\bf r}_2) \quad .
\end{equation}
A combined index $L$ is used 
for the specification of the symmetry. In this work it is limited to   
either $^1\Sigma_g$ ($L=0$) or $^1\Sigma_u$ ($L=1$) symmetry. 
The index $n=1,2,\dots$ just numbers the different states for a 
given symmetry $L$. 

The field-free two-electron wavefunctions $\phi_{nL}$ are 
obtained by a configuration-interaction (CI) calculation performed 
in a basis of {\htpl} orbitals. The {\htpl} orbitals are
expressed in a $B$-spline basis set in prolate-spheroidal coordinates. 
Since the basis functions are confined to a finite spatial volume, 
a discretized representation of the electronic continuum is obtained.
Substitution of Eq.~(\ref{eq:expansionpsi}) into Eq.~(\ref{eq:fulltdse})  
yields then a finite set of coupled differential equations for the 
time-dependent coefficients $b_{nL}$. These equations are solved 
numerically with the initial condition 
$b_{nL}(t=0)\,=\,\delta_{n,1}\,\delta_{L,1}$ for the {\bsig}  state.

In the present work, the TISE for {\htpl} is solved in a box of size
350\,$a_0$. The solution of the TISE for {\htpl} is obtained 
in prolate spheroidal coordinates 
($1 \, \leq \, \xi \, < \, \infty, -1 \, \leq \, \eta \, \leq \,
1, 0\,\leq\,\phi\,<\,2\pi$). The total symmetry of the {\htpl} states is
defined on the basis of angular momentum and {\em gerade} or {\em ungerade}
symmetry of the state wavefunction. For each total symmetry of {\htpl} 350 $B$
splines of order 15 are used in $\xi$ direction and 24 $B$ splines of order 8
in $\eta$ direction. {\htpl} states with angular momenta between 0 and 5 
with {\em gerade} and {\em ungerade} symmetry are calculated. Using such a 
basis set yields around 4200 states for each total symmetry of {\htpl}. The CI
calculation performed with these {\htpl} states gives around 8000 states for
each symmetry of {\hmol} ($^1\Sigma_g^{+}$ and $^1\Sigma_u^{+}$).

For computational convenience, cos$^2$-shaped pulses%
\begin{equation}\label{eq:cos2}
  X(t) \, = \, X_0\, \cos^2(\pi\,t/\tau) \, \cos(\omega\,t \, 
                          + \, \phi_{\rm CEP})
\end{equation}
where $\omega$ is the photon energy and $X$ stands for $F$ (length form) or
$A$ (velocity form) are used in most of the calculations of this work.  (Note,
the pulses defined this way are not identical for $X=F$ or $X=A$, especially
in the case of extremely short pulses.)  The peak electric field or vector
potential amplitudes are $F_0$ and $A_0$, respectively. In order to obtain the
maximum field at the center of the pulse, the carrier-envelope phase (CEP)
$\phi_{\rm CEP}$ is set to 0 ($\pi/2$) for the length (velocity) form.  The
advantage of the cos$^2$ pulses is the finite pulse length $\tau$ (in the
interval $[-\tau/2 , \tau/2]$) and thus the well-defined interval for time
integration. This is an evident advantage compared with the (more realistic)
Gaussian pulse
\begin{equation}\label{eq:gaussian}
  X(t) \, = \, X_0\,\exp\left(-\frac{2\,t^2}{{\tau_g}^2}\right) \, 
                 \cos(\omega \, t \, + \,\phi_{\rm CEP} )
\end{equation}
that only exponentially decays to zero and requires thus a careful convergence
study with respect to the integration interval. In order to assure that found
deviations to the calculation in \cite{sfm:bart08} are not caused by a
possible difference in the definition of the pulse shape, the results of
cos$^2$ and Gaussian pulses are compared. In this case the characteristic time
$\tau_g$ of the Gaussian pulse is chosen according to%
\begin{equation}\label{eq:GaussCos2}
   \tau_g \, = \, \frac{\tau}{4} \, \sqrt{-\frac{1}{2}\ln\left(\frac{1}{2}\right)}\, \quad .
\end{equation}
This choice yields the same full width at half maximum (FWHM) of the 
Gaussian and the cos$^2$ pulses. 
 
For sufficiently low intensities one expects lowest-order perturbation 
theory (LOPT) to adequately describe the interaction of a molecule 
with a laser field. Within LOPT the $N$-photon ionization rate 
$\Gamma^{(N)}$ (in s$^{-1}$) is given by
\begin{equation}\label{eq:loptrate}
 \Gamma^{(N)} \, = \, \sigma^{(N)} \, \left( \frac{I}{\hbar \omega} \right)^N
\end{equation}
where $I$ is the intensity of the monochromatic laser field in {\wcm}, $\hbar
\omega$ is the photon energy in Joule, and $\sigma^{(N)}$ is the generalized
$N$-photon ionization cross section in cm$^{2N}$s$^{N-1}$. 
For single-photon ionization $\sigma^{(1)}$ is equal to the standard one-photon 
cross-section and LOPT reduces to Fermi's Golden Rule.  
Integration of Eq.~(\ref{eq:loptrate}) over the intensity profile of a pulse
gives the ionization yield
\begin{equation}\label{eq:loptyield}
  P_{\rm ion}^{\rm LOPT} \, = \, 
           1 - \exp \left( - \int_{\rm Pulse} \Gamma^{(N)}(I(t))\; dt 
                    \right) 
\end{equation}
where $\Gamma^{(N)}$ is a function of intensity $I$ 
(see Eq.\,(\ref{eq:loptrate})) which for a laser pulse is a function of 
time $t$. The generalized cross-sections $\sigma^{(N)}$ 
are in this work calculated with the same field-free two-electron 
states $\phi_{nL}$ of Eq.\,(\ref{eq:h2tise}) that are used for solving 
the TDSE. Results within LOPT have been earlier obtained for H$_2$ 
in \cite{sfm:apal01, sfm:apal02} and a comparison to TDSE results was given in 
\cite{sfm:awas05}. In those references technical details on the 
evaluation of $\sigma^{(N)}$ may be found. However, in those previous 
investigations only ionization from the electronic ground state was 
considered. 

For a more detailed investigation of the validity of the SAE approach
introduced in \cite{sfm:bart08} it is of interest to consider also alternative
SAE models. In one approach the TDSE is solved as discussed above, but a
restricted set of configurations is used in the CI calculation. In this case
only configurations are adopted in which one electron occupies the lowest
lying ($1\,\sigma_g$) orbital of {\htpl}.  The other electron may then occupy
any of the $n\,\sigma_g$ ($n\,\sigma_u$) orbitals to yield the configurations
for the CI calculation giving the $^1\Sigma_g$ ($^1\Sigma_u$) states of H$_2$.
This approach is referred to as pseudo SAE (p-SAE)
\cite{sfa:lamb98a,sfm:awas08,sfa:nubb08}. The p-SAE approach corresponds to a 
complete relaxation of the core. A frozen-core SAE approach where there is no
relaxation of the core has recently been introduced in \cite{sfm:awas08}. In
that case the core electrons are frozen in their field-free initial-state
orbitals and the TDSE is solved for the active electron in the combined field
of the core and the laser pulse. In the implementation the core was described
within density-functional theory (DFT). For strong-field excitation and
ionization of ground-state H$_2$ it was, however, demonstrated that the
results agree quite well with the ones obtained with a core described within
Hartree-Fock theory, despite the rather different electronic binding energies.
The DFT-based variant used in this work will be referred to as DFT-SAE.

Two of the SAE approaches, p-SAE and DFT-SAE, give the ionization yield for a
single electron. In \cite{sfm:awas08} it was found that for ionization yields
of up to about 10\,\% the SAE results for H$_2$ should be multiplied by a
factor 2 in order to properly account for the two equivalent
electrons. However, for larger ion yields this factor 2 leads to an
overestimation, because the screening of the core electron is reduced. As a
consequence, the ionization potential increases and the ionization probability
decreases. Therefore, the ion yield of the SAE approximation approaches the
full CI-TDSE result for high intensities and ion yields larger than about
20\,\%.  However, this screening argument applies in principle only, if the
ionization process depends strongly on the electronic binding energy. This is,
e.\,g., the case for ionization in the tunneling picture, but is not expected
to be valid in the opposite extreme of perturbative single-photon
ionization. It should be noted that no prefactor should be used in the
Koopmans'-picture based SAE approach (K-SAE) of 
Barth {\em et al.}~\cite{sfm:bart08}.

\section{Results}\label{sec:results}
%
%
%
%
\begin{table*}
  \caption{\label{tab:energy_R2p3} State energies (in atomic units E$_h$)
    using different approaches at $R=2.2998$\,a$_0$. The ionization potential
    of the {\bsig} states is shown in the last column.}
\begin{center}
\begin{ruledtabular}
\begin{tabular}{l l l l l}%
Approach & {\hmol}\, {\xsig} & {\hmol}\, {\bsig}   &
{\iong} & I$_{\rm p}$({\bsig}) \\ \hline
CI      & $-1.10374$ &  $-0.7485$ & $-0.5989$  &  $0.1496$  \\ 
p-SAE   & $-1.06385$ &  $-0.7388$ & $-0.5989$  &  $0.1399$  \\ 
DFT-SAE & $-1.08196$ &  $-0.8381$ & $-0.5989$  &  $0.2392$  \\ 
K-SAE \cite{sfm:bart08}
        &  n.\,a.    &  $-0.7509$ & $-0.5988$  &  $0.1521$  \\ 
Accurate  & $-1.1111725$~\cite{dia:kolo65} & $-0.7563608186$~\cite{dia:woln88}
& $-0.5989$ & $0.1575$ \\ 
\end{tabular}
\end{ruledtabular}
\end{center}
\end{table*}

%
In Table~\ref{tab:energy_R2p3} the energies of the ground ({\xsig}) and the
first excited ({\bsig}) states of {\hmol} are given at the internuclear
separation $R=2.2998$\,a$_0\,$. Furthermore, the corresponding ground-state
energies of the {\htpl} ion are shown together with the ionization potential
of the {\bsig} state that follows from them. The energies obtained by means of
K-SAE (given in \cite{sfm:bart08}), CI, p-SAE, and DFT-SAE approaches are
compared.  The best available theoretical energy values are also shown. 
While the energy of the {\bsig} state obtained with K-SAE is in very 
good agreement with the CI calculation, the 
DFT-SAE result is almost 0.09\,a.u.\ (2.4\,eV) away. The
p-SAE result is much closer to the CI result, but not as close as K-SAE. Since
the ionic ground-state energies agree very well for all approaches, the
differences found for the ionization potentials are a result of the different
{\bsig} energies. For completeness, also the different ground-state energies
are given, if available. In this case, DFT-SAE agrees better to CI than
p-SAE. It was demonstrated in \cite{dia:vann04,dia:vann06} that very accurate
CI results can be obtained with the present approach, if the basis-set
parameters including the configurations are judiciously chosen. However, for
solving the TDSE a compromise has to be made, since a large spectrum of
field-free states including the electronic continuum has to be described with
reasonable accuracy.

The reason for the very poor DFT energy of the {\bsig} state is probably
twofold. First, small systems like He and H$_2$ are known to be difficult to
be described by DFT. More importantly, however, standard DFT is in principle
not applicable to excited states. In practice, time-dependent DFT is
usually adopted to obtain excitation energies.  Nevertheless, it should be
interesting to investigate the performance of the recently developed DFT based
SAE in comparison to the also very newly proposed K-SAE as is done below. 

\subsection{8\,eV photons}\label{ss:om8ev}
In \cite{sfm:bart08} it was concluded that single-photon ionization of the
{\bsig} state of H$_2$ at 8\,eV should satisfy the conditions for the
applicability of a time-dependent extension of Koopmans' picture, a
single-active-electron approximation based on Koopmans' theorem, i.\,e.\
within K-SAE. Therefore, the response to a 5-cycle laser pulse with 8\,eV
photons was studied in the framework of the newly developed K-SAE. The TDSE
was solved on a grid using the length formulation. In \cite{sfm:bart08}
the calculated ionization yield plotted against the laser peak intensity on
a log-log scale showed a slope 1 in a large range of intensities. This
confirmed the perturbative nature of the single-photon process in the selected
intensity window and was interpreted as a further confirmation of the proper
implementation of the time-propagation in the K-SAE approach. Note, in
\cite{sfm:bart08} the laser peak intensity is defined as $I = \epsilon_0 c
F_{0}^{2}$ where $\epsilon_0$ is the vacuum permittivity, $c$ is the vacuum
speed of light, and $F_0$ is the peak value of the electric field. In this
work the alternative, cycle averaged peak intensity $I = \frac{1}{2}
\epsilon_0 c F_{0}^{2} $ is used.

\begin{table}[ht]
\caption{\label{tab:saeresults8ev} Comparison of K-SAE, CI, LOPT, p-SAE, and
  DFT-SAE results. The populations of the initial {\bsig} state and the
  ionization yields are given for two different values of the peak electric
  field. The K-SAE results are taken from \cite{sfm:bart08}.}
\begin{center}
\begin{ruledtabular}
\begin{tabular}{lllllll}%
 & F$_0$ (a.u.) & K-SAE & CI  & LOPT & p-SAE & DFT-SAE   \\ \hline
Initial & 0.005 & n.a.   & 0.960 & 0.963  & 0.967 &   0.980 \\ 
 Ion.    & 0.005 & 0.0043 & 0.033 & 0.037 & 0.0031 &  0.022  \\ 
Initial & 0.020 & 0.930 & 0.548 & 0.543  & 0.624 & 0.728 \\ 
Ion.    & 0.020 & 0.066 & 0.379 & 0.457 & 0.036 & 0.261 \\
\end{tabular}
\end{ruledtabular}
\end{center}
\end{table}

\begin{figure}[ht]
\centering\includegraphics[width=0.9\linewidth]{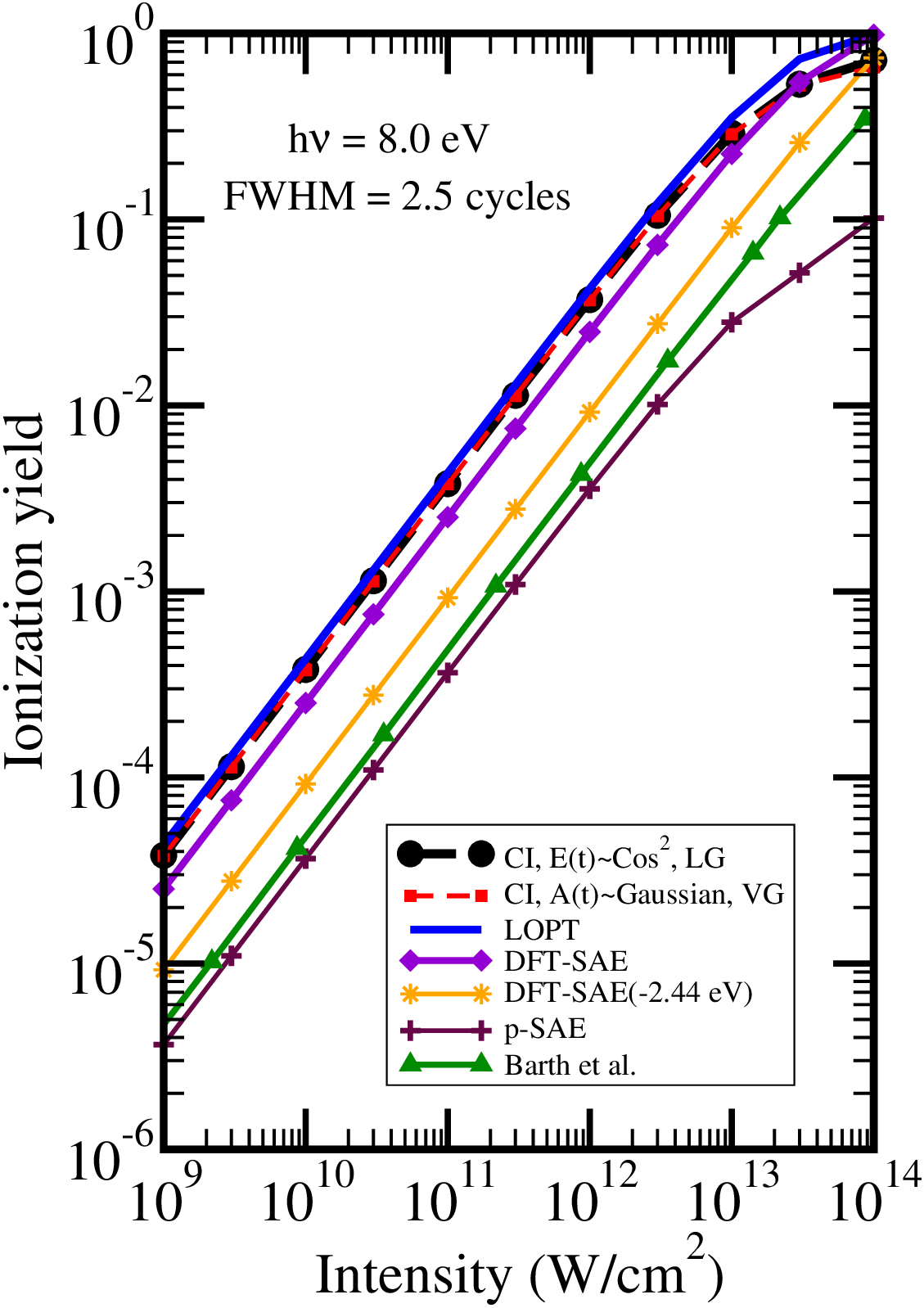}
\caption{Ionization yield as a function of the peak laser intensity. The TDSE
  calculation is performed using length-form transition dipole moments with a
  $\cos^2$ envelope for the electric field (black, circles) and velocity-form
  transition dipole moments with a Gaussian envelope for the vector potential
  (red, squares). The FWHM for both pulses is $2.5\,$ cycles of $8\,$eV photon
  energy. The ionization yields obtained with LOPT (blue, solid line), DFT-SAE
  (violet, diamonds), scaled DFT-SAE by shifting the photon energy by 
  $-2.44\,$eV (at 10.44\,eV) (orange, asterix), and p-SAE (maroon, plus)
  are also shown and compared to the results from Barth 
  {\it et al.}~\cite{sfm:bart08} (green, triangles). }
\label{fig:H2_R2p3_8ev_FWHM2p5_ion}
\end{figure}

Besides the graphically shown intensity scan, the ionization yield and (in one
case) the population remaining in the initial {\bsig} state of H$_2$ were
given in numerical form for the peak electric fields $F_0=0.005$ and 
$0.02\,$a.\,u.\ in \cite{sfm:bart08}. In Table~\ref{tab:saeresults8ev} these 
results are compared to the corresponding values obtained in this work within 
the different approaches, CI-TDSE, LOPT, p-SAE, and DFT-SAE. In LOPT the
initial-state population $P_{\rm ini}$ was defined as $P_{\rm ini} = 1-P_{\rm
  ion}^{\rm LOPT}$.

The ionization yield predicted on the basis of K-SAE differs substantially
from the CI-TDSE result. At the lower field strength ($F_0=0.005\,$a.\,u.)
the K-SAE yield is almost one order of magnitude (factor 7.7) too small.  For
$F_0=0.02\,$a.\,u.\ the disagreement is a little bit smaller (factor
5.7). While p-SAE deviates from the CI-TDSE results even more than K-SAE
(underestimation by factors 10.6 and 10.5), DFT-SAE appears on the first
glance to be the SAE model with the
best agreement to the full two-electron result, despite the poor energy of the
{\bsig} state.  Although the DFT-SAE yields are also smaller than the CI-TDSE
ones, they differ only by factors 1.5 and 1.45 for the smaller and the larger
field strengths, respectively. Clearly the best agreement to the CI-TDSE
result is, however, obtained with LOPT. At the lower field strength the 
agreement is in fact good (the CI-TDSE yield is overestimated by a factor 
1.1 only) and for the larger field strength it is still reasonable (factor
1.2). Note, the given fields correspond to laser peak intensities of about 
$8.8 \times 10^{11}\,${\wcm} and $1.4 \times 10^{13}\,${\wcm} where LOPT is not
necessarily expected to work. The one-photon cross-section at 8\,eV is found
to be $\sigma^{(1)} = 5.75 \times 10^{-17}\,$cm$^2$.

A Taylor expansion of the exponential in Eq.\,(\ref{eq:loptyield}) and insertion
of Eq.\,(\ref{eq:loptrate}) shows that for one-photon ionization the ion
yield is proportional to the one-photon absorption cross-section and the laser
intensity, $P_{\rm ion}^{\rm LOPT} \propto \sigma^{(1)}\,I$, if the time 
integral is small. Therefore, one expects within LOPT and for
small ion yields that an increase of the field strength by a factor 4 leads
to an increase of the ion yield by a factor $4^2=16$. This factor is
reproduced well for K-SAE (15.3), but neither for CI-TDSE (11.5) nor for LOPT
(12.4). The latter result shows that for the larger ionization yields found
for the two-electron models CI-TDSE and LOPT the assumption of a sufficiently
small time integral is not fulfilled and the ion yield does not increase
linearly with intensity, even within LOPT. Correcting the K-SAE yield at the
larger field for this effect shows that the agreement between K-SAE and
CI-TDSE does in fact not improve for the higher field, but the disagreement
remains at a factor of about 7.5. While the factor 11.9 found between the two
ion yields obtained with DFT-SAE appears understandable from the similar
magnitude of ionization as compared with CI-TDSE and LOPT, one would expect
for p-SAE a factor close to 16 similar to K-SAE, since the ion yield is even
smaller. However, instead a value 11.6 is found, similar to the other cases
except K-SAE.

The pronounced failure of K-SAE despite the very accurate energy of the
{\bsig} state is clearly surprising. In order to clarify this issue it was
checked that it is not an artifact of the different numerical
implementations. One possibility could, e.\,g., be a different pulse
definition, or the use of the velocity or length forms of the dipole operator.
In Fig.~\ref{fig:H2_R2p3_8ev_FWHM2p5_ion} the ionization yield as a function
of the laser peak intensity is shown. It confirms the already discussed good
agreement of CI-TDSE with LOPT. The ionization yield obtained by means of
DFT-SAE is quite close to the full CI-TDSE results. The p-SAE and K-SAE fail
by almost an order of magnitude for 8\,eV photons. In order to investigate the
sensitivity of the results on pulse details and the form of the dipole
operator used, two CI-TDSE results are shown in
Fig.~\ref{fig:H2_R2p3_8ev_FWHM2p5_ion}. In one case a cos$^2$ pulse envelope  
(for $E(t)$) is used with dipole moments in length form, while in the 
other case a Gaussian pulse shape (for $A(t)$) is used together with 
the velocity form. Clearly, neither the choice of
the dipole operator nor the exact pulse shape (in both cases a FWHM of 2.5
cycles is used) modifies the result substantially. In fact, the agreement is
almost perfect, except for very high intensities and thus very close to
saturation.

In view of the rather poor energy of the {\bsig} state in DFT-SAE the superior 
result compared to the other SAE models is suspicious. In fact, as has been 
discussed previously in \cite{sfa:lamb00,sfm:apal02,sfm:awas05} a poor
initial-state energy should be accounted for by a correspondingly shifted 
photon energy. This energy shift should be chosen in such a way that the 
correct multi-photon ionization threshold (in this case the one-photon 
threshold) is obtained. Since the present DFT threshold lies about 2.44\,eV 
higher than the one of the CI calculation (see Table~\ref{tab:energy_R2p3}), 
the CI-TDSE results for a photon energy of 8\,eV should be compared to the 
DFT-SAE results at a photon energy of 10.44\,eV. The corresponding curve 
(also drawn in Fig.~\ref{fig:H2_R2p3_8ev_FWHM2p5_ion}) shows that the 
agreement with CI-TDSE worsens evidently, although it remains better 
than the agreement of the other SAE models with the CI-TDSE result. 
It is clear, however, that the (slightly) better agreement of DFT-SAE 
appears to be accidental.
   
It is, of course, interesting to investigate the origin of the rather evident 
failures of K-SAE and p-SAE.  A characteristic property of the {\bsig} state 
of H$_2$ is its rather large ionic component \cite{dia:shar71}. In fact, it 
was shown theoretically that this ionic character plays a crucial role in 
the strong-field behaviour of H$_2$. For example, it is responsible for bond 
softening and enhanced ionization of {\it neutral} 
H$_2$ \cite{sfm:saen00a,sfm:saen00b,sfm:haru00,sfm:kawa00,sfm:saen02a}.  
The field-induced admixture of the ionic (H$^+$H$^-$) component to the {\xsig} 
ground state leads to strongly enhanced ionization, since H$^-$ possesses a 
very low electronic binding energy.  It appears quite reasonable that SAE 
models fail to properly describe the ionic component of the {\bsig} state, 
since the (frozen) spectator electron always occupies a molecular orbital 
that is symmetrically distributed over both protons. As already discussed, 
the lack of even a very small admixture of the ionic component may reduce the 
ionization rate in a strong electric field significantly. This could explain 
the too low ionization yield found uniformly for all SAE models discussed 
in this work. 

However, the good agreement between LOPT and CI-TDSE (especially at low
intensities) allows to conclude that one should not look for a strong-field
explanation where the binding energy influences the ionization rate in an
exponential way. Instead, the validity of the LOPT indicates that it is only
the (direct) transition dipole matrix element between the initial {\bsig}
state and the continuum (reached with an 8\,eV photon) that determines the
magnitude of ionization. In fact, a rough estimate based on a simplified model
describing ionization from such an ionic pair state (weighted by the
contribution of this state to the {\bsig} state of H$_2$) indicates that this
ionic component cannot at all explain a massive increase of the one-photon
ionization yield. Another possible explanation of the failure of a SAE model
could be the occurrence of doubly-excited autoionizing states that are
necessarily absent in SAE models. However, they are expected to be excited at
photon energies larger than 8\,eV~\cite{dia:sanc97b, dia:vann06}.

\subsection{One-photon ionization}\label{ss:sinphoion}

\begin{figure}[ht]
\includegraphics[width=0.95\linewidth]{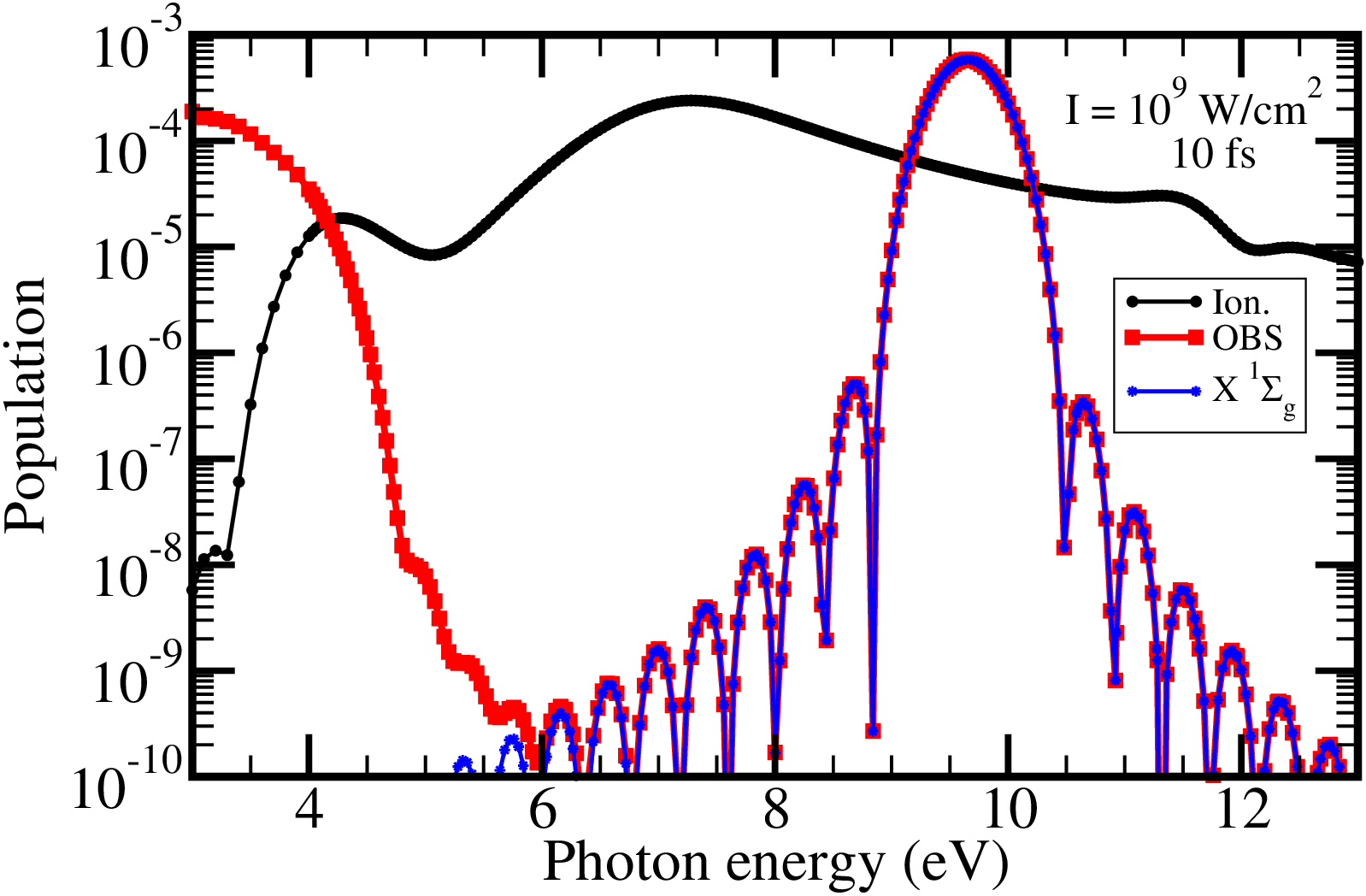}
\caption{Ionization (black, circles), population of the other bound states OBS
  (red, dashed), and population of the {\xsig} ground state (blue, asterix)
  obtained with CI-TDSE as a function of the incident photon energy for a
  $10\,$fs laser pulse with a peak intensity of $1.0 \times 10^{9}\,${\wcm}.}
\label{fig:H2_R2p3_m007_SPI}
\end{figure}

A more complete picture of one-photon ionization from the {\bsig} state of
H$_2$ may be obtained from a consideration of the photon-energy dependence.
Figure~\ref{fig:H2_R2p3_m007_SPI} shows the ionization yield of the H$_2$
{\bsig} state exposed to a $10\,$fs laser pulse with a peak intensity of $1.0
\times 10^{9}\,${\wcm} obtained with CI-TDSE. The photon energy varies between
about 3 and 13\,eV and covers thus the one-photon regime from just below the
ionization threshold to the energy at which the first excited electronic state
of H$_2^+$ becomes energetically accessible. Due to the finite pulse width
(and possible two-photon ionization) the ionization starts before the
threshold energy of about 4.1\,eV. In contrast to one-photon ionization
starting from the electronic ground state which shows an overall monotonous
decrease with increasing photon energy~\cite{dia:sanc97a}, the ionization
yield of the {\bsig} state decreases until about 5\,eV after which it
increases drastically (note the logarithmic scale) until about 7\,eV.  Beyond
about 7\,eV the yield decreases monotonously, until some structure due to the
occurrence of autoionizing states becomes visible in between about 11\,eV and
12.5\,eV. Clearly, the autoioinizing states cannot explain the failure of the
SAE approaches at 8\,eV discussed in Sec.~\ref{ss:om8ev}.

The population of the H$_2$ {\xsig} ground state after the laser pulse 
is also shown in Figure~\ref{fig:H2_R2p3_m007_SPI}. A pronounced maximum
occurs at the expected energy of about 9.66\,eV at which the {\bsig}
$\rightarrow$ {\xsig} transition becomes resonant. The width of the peak
reflects the spectral width of the laser pulse. For the considered
laser peak intensity the population of the ground state and thus the
probability for deexcitation of the {\bsig} state is at its maximum about 
a factor 10 larger than the ionization yield. Since the total depopulation 
of the {\bsig} state is, however, still very small (less than 0.1\,\%), 
the ionization probability remains practically uninfluenced by this 
deexcitation process. As was shown recently in 
\cite{sfm:vann08}, the oscillations (side peaks) of the resonant peak are 
due to the cos$^2$ shape of the pulse used in the calculation. Since they 
are many orders of magnitude smaller than the ionization yield, the latter 
is, however, not influenced by these spurious oscillations. This is also 
confirmed explicitly for 8\,eV photons in 
Fig.~\ref{fig:H2_R2p3_8ev_FWHM2p5_ion}, since the ion yields obtained with 
a Gaussian pulse that does not show these oscillations agrees well with 
the ones obtained with a cos$^2$ pulse. 

The population of the other bound states (OBS) is also shown in
Figure~\ref{fig:H2_R2p3_m007_SPI}. It is defined as 
$P_{\rm OBS} = 1 - ( P_{\rm ini} + P_{\rm ion})$
where $P_{\rm ini}$ is the population of the initial {\bsig} state and $P_{\rm
  ion}$ is the ionization yield). In a large range of photon energies
$P_{\rm OBS}$ is dominated by the ground-state contribution. Only below 6\,eV 
the energetically higher lying bound states come into play. Close to the 
ionization threshold the ion yield and $P_{\rm OBS}$ cross each other. 
Below threshold the Rydberg states of H$_2$ are resonantly populated, while 
the one-photon ionization channel closes. As already mentioned, the finite 
band width of the adopted laser pulse leads to a rather broad energy 
range in which excitation and ionization compete with each other.

\begin{figure}[ht]
\includegraphics[width=0.95\linewidth]{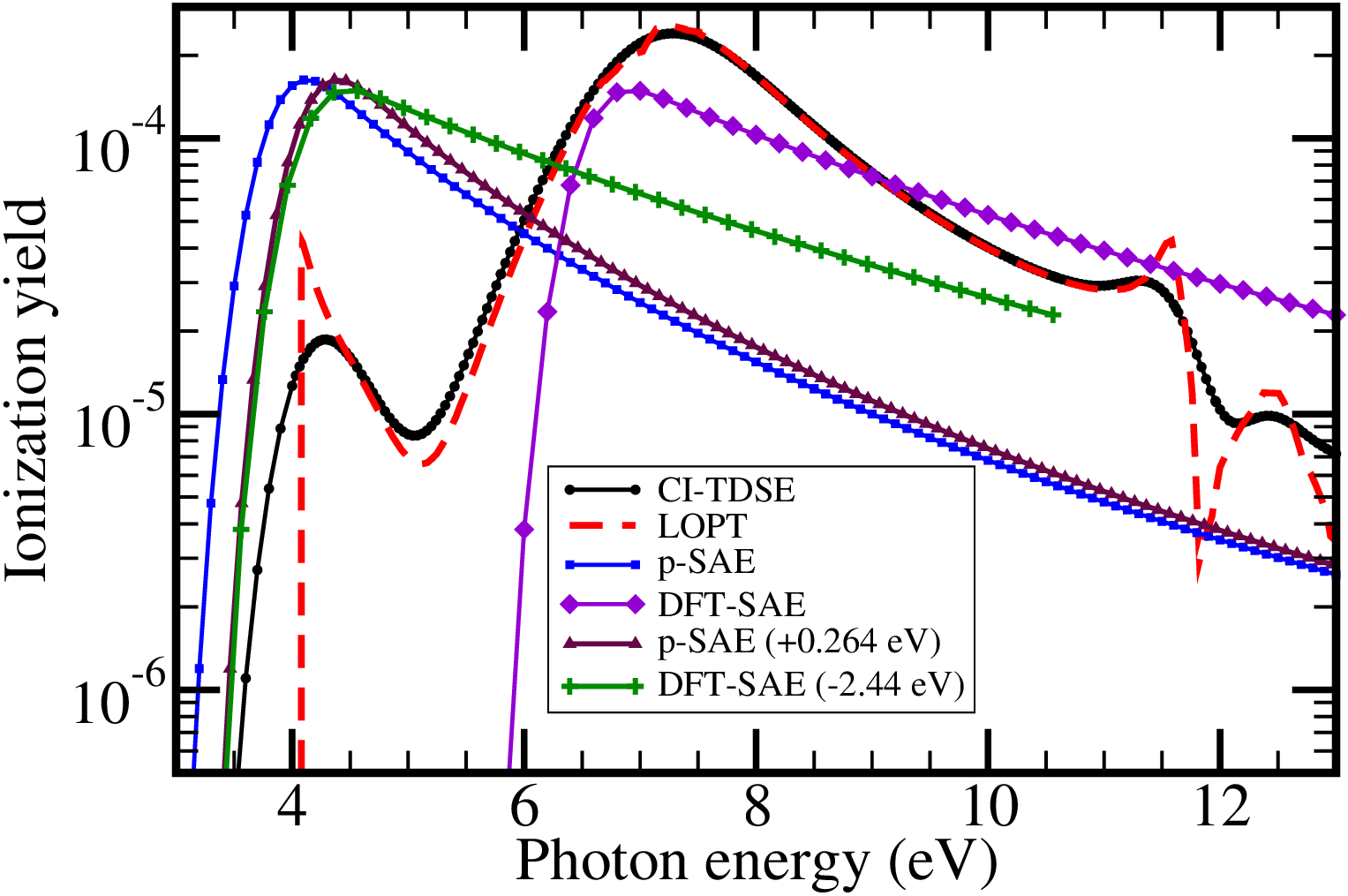}
\caption{Photon-energy dependent ionization yield for various theoretical
  models with laser parameters as in Fig.~\,\ref{fig:H2_R2p3_m007_SPI}.  The
  ionization yields obtained with CI-TDSE (black, circles), LOPT(red, dashed),
  p-SAE (blue, square), and DFT-SAE (violet, diamonds) are compared. The
  ionization yields obtained by a shift of the photon energy to compensate the
  corresponding error in the ionization potential with respect to CI of p-SAE
  (maroon, triangles) and DFT-SAE (green, plus) are also shown.}
\label{fig:H2_R2p3_SPI_SAE_ion_compare}
\end{figure}

Figure~\ref{fig:H2_R2p3_SPI_SAE_ion_compare} shows the ionization yields
obtained by means of the different theoretical models in the one-photon
ionization regime. The laser parameters are the same as the ones in
Figure~\ref{fig:H2_R2p3_m007_SPI}. The ionization yield obtained by CI-TDSE as
shown in Figure~\ref{fig:H2_R2p3_m007_SPI} is repeated for comparison.  For
the chosen laser intensity the Keldysh parameter is much larger than 1 even at
the lowest photon energy.  This indicates ionization to take place in the
multiphoton regime (although the name is evidently a bit misleading in the
present case of one-photon ionization). In fact, as the comparison to the
prediction of LOPT shows, the process is quite well described by perturbation
theory in almost the complete energy range. The observed deviations close to
the ionization threshold and in the regime of autoionizing states are mainly
due to the finite pulse width in the TDSE calculation. Since LOPT predicts
ionization rates, an infinite pulse duration is implied. As a consequence, a
sharp ionization threshold and sharp resonances with the widths of the latter
determined only by the corresponding lifetimes are observed in the LOPT
spectrum. A more appropriate comparison would thus involve to convolute the
LOPT spectrum with the spectral band width or to consider longer pulses in the
TDSE calculation. In fact, it is interesting to note that already for a pulse
as short as 10\,fs such a good agreement is found. For the present discussion
the most important conclusion is, however, the very good agreement of LOPT and
CI-TDSE in the non-resonant photon-energy range between about 6 and 10.5\,eV
in which sharp features are absent.

The ionization yield obtained by solving the TDSE in the p-SAE approximation 
shows beyond the ionization threshold a monotonous decreasing behavior with 
increasing photon energy. This is very similar to the behavior found for 
one-photon ionization from the electronic ground state. In order to correct 
for the inaccurate ionization potential obtained within the p-SAE approach 
the ionization yield may be shifted by 0.264\,eV in order to compensate this 
effect. However, even after the shift the ion yields obtained with the
p-SAE-TDSE approach or CI-TDSE differ substantially. At the threshold 
the p-SAE result is about one order of magnitude larger than the CI 
yield. At about 6\,eV the ion yields of the p-SAE and CI approaches 
cross each other. At the maximum of the CI ionization yield close to 
7.3\,eV the CI result is about an order of magnitude large than the 
p-SAE result. Apart from the local variation in the resonant regime, 
the ratio between CI and p-SAE results decreases then rather monotonously 
to about a factor 2.5 at 13\,eV. Clearly, p-SAE is completely inadequate 
for predicting the one-photon ionization yield of the {\bsig} state 
of H$_2$ in the whole considered photon-energy range, and not only at 
8\,eV.  

The DFT-SAE ionization yield shifted by $-2.44\,$eV in order to compensate the
wrong ionization potential agrees at the threshold, i.\,e.\ between about
3.5\,eV and 4.75\,eV quite well with the p-SAE result. However, although the
DFT curve decreases also monotonously with increasing photon energy, its
decrease is much slower compared to the one predicted within the p-SAE
approach. As a consequence of the smaller slope, the quantitative deviation
between DFT-SAE and the CI calculation is smaller than the one between p-SAE
and CI. In view of the massive qualitative difference between the DFT-SAE and
the CI curve this could, however, be accidental. Clearly, the reasonable
agreement between the {\it unshifted} DFT-SAE and the CI result at 8\,eV is
definitely accidental. In conclusion,
Fig.~\ref{fig:H2_R2p3_SPI_SAE_ion_compare} shows that both SAE approaches
(p-SAE and DFT-SAE) fail completely when describing one-photon ionization of
the {\bsig} state of H$_2$. Quantitatively, there is a disagreement of up to
an order of magnitude. Due to the completely different qualitative behavior,
the SAE results may under- or overestimate the true ionization yield. The
result of the K-SAE at 8\,eV \cite{sfm:bart08} seems to indicate, that a
similar failure as is found for p-SAE is to be expected also for the K-SAE,
despite the relatively accurate initial-state description of the latter. In
fact, as is shown in the next section, it is really a failure of the SAE
approaches to properly describe the final continuum states.

\subsection{Explanation for the failure of the SAE approximation}\label{ss:explain}

\begin{figure}[ht!]
\includegraphics[width=0.95\linewidth]{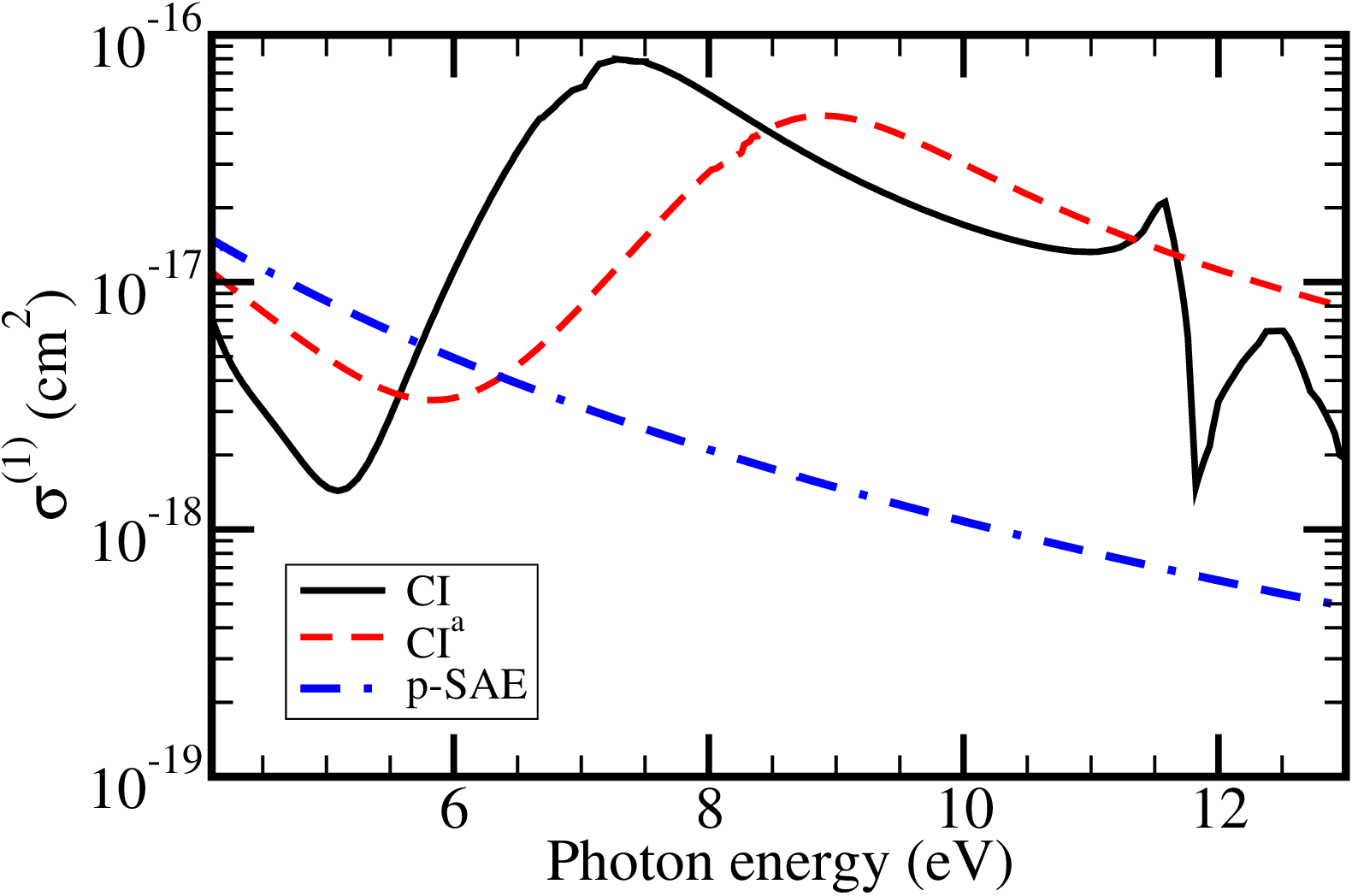}
\caption{The one-photon ionization cross-section (within LOPT) of the H$_2$
  {\bsig} state for the CI (solid, black), the CI$^{\rm a}$ (red, dashed), and
  the p-SAE (blue, chain) basis sets.} 
\label{fig:H2_R2p3_SPI_CRS}
\end{figure}

In order to analyze the origin of the massive deviation between the 
SAE and the CI ionization yields it may be noted that the p-SAE approach 
differs from the full CI calculation only by a restriction of the 
configurations that are included in the CI calculation, 
as was outlined in Sec.~\ref{sec:method}. Furthermore, the previous 
results have demonstrated that LOPT is applicable and the deviations 
between the CI and the p-SAE LOPT cross sections is practically identical 
to the ones found between CI-TDSE and p-SAE-TDSE.  However, LOPT allows 
a much easier analysis of the failure of the SAE approximation, as is 
outlined below.

The origin of the failure of the SAE approaches can be understood from 
Fig.~\ref{fig:H2_R2p3_SPI_CRS} that shows the one-photon ionization 
cross-sections within LOPT obtained from CI calculations with 
3 different basis sets. The CI basis that gave converged results is 
again denoted as CI. As already explained, keeping in the CI basis only 
those configurations in which one electron remains in the 1\,$\sigma_g$ 
ground-state orbital of H$_2^+$ leads to the p-SAE model. 
The CI$^{\rm a}$ basis set is almost identical to the 
p-SAE basis set. The only difference is that it includes one additional 
{\susu} configuration in the CI calculation for the states with 
$^1\Sigma_g^+$ symmetry. Thus the initial {\bsig} state is identical 
for the CI$^{\rm a}$ and the p-SAE basis sets. Only the final continuum 
states differ by the additional admixture of the single configuration 
({\susu}). 

As is seen from Fig.~\ref{fig:H2_R2p3_SPI_CRS}, the addition 
of this configuration to the CI calculation of the final states leads to 
a substantial modification of the ionization cross-section. While for 
photon energies below about 6.5\,eV the cross-section obtained 
with the CI$^{\rm a}$ is reduced compared to the p-SAE result, 
there is a pronounced enhancement by more than an order of magnitude 
at higher photon energies. While there is some evident deviation between 
the full CI calculation and the CI$^{\rm a}$ results, it is clear that 
the main failure of the SAE approximation is due to the lack of the 
{\susu} configuration in the description of the final continuum states. 
A further inclusion of configurations is required for achieving 
quantitative convergence, but the cross section does not change 
qualitatively. The reduction at small energies becomes more pronounced 
and both minima and maxima are shifted to lower photon energies. 
Clearly, also the adequate description of the spectral features 
above 11\,eV due to autoionizing states requires the inclusion of 
additional configurations. 

Based on this finding a simple semi-quantitative model for the 
failure of the SAE approximation can be developed. The one-photon 
ionization cross-section is proportional to $|\langle \psi_f |
\op{D} | \psi_i \rangle|^2$, where $\psi_f$ and $\psi_i$ are the final 
and the initial state wavefunctions, respectively. $\op{D}$ is in this 
case the time-independent electronic dipole moment operator. In the 
simplest SAE approximation in which one electron is 
frozen during the whole process the initial state 
{\bsig} can be described as a product of the (mean-field) wavefunctions 
$| 1\sigma_g \rangle$ and $| 1\sigma_u \rangle$ of H$_2$. 
A more sophisticated SAE model accounts for the proper (anti-)symmetry 
of the electronic wavefunction (as in the p-SAE model), but this does 
not change any of the following conclusions as can easily be verified. 
In order to keep the notation simple, the simple frozen-core model 
is thus pursued. The final state $\psi_f$ can then be 
approximated as $| 1\sigma_g \rangle$ $| \epsilon\sigma_g \rangle$ 
in this simple SAE approximation and as 
$C_1 (| 1\sigma_g \rangle$ $| \epsilon\sigma_g \rangle) \,
+\, C_2 (|1\sigma_u \rangle |1\sigma_u \rangle)$ in a simplified 
two-electron model with (some) correlation. 
Here, $\epsilon\sigma_g$ describes the corresponding continuum 
orbital with energy $\epsilon$ and one has $|C_1|^2\,+\,|C_2|^2\,=\,1$. 
The dipole transition matrix element in the SAE case is
\begin{equation}\label{eq:DIPSAE}
  \langle \psi_f | \op{D} | \psi_i \rangle = \langle 1\sigma_g ; 
  \epsilon\sigma_g | \op{D} | 1\sigma_g ; 1\sigma_u \rangle = 
  \langle \epsilon\sigma_g | \op{D} | 1\sigma_u  \rangle \,
\end{equation}
while the two-electron model yields  
\begin{equation}\label{eq:DIPCIA}
\langle \psi_f |\op{D}| \psi_i \rangle = 
     C_1 \langle \epsilon\sigma_g | \op{D} | 1\sigma_u  \rangle  
   + C_2 \langle 1\sigma_u | \op{D} | 1\sigma_g  \rangle \, .
\end{equation}
Evidently, the latter gets an additional contribution from the doubly 
excited $|1\sigma_u \rangle |1\sigma_u \rangle $ configuration. 

Of course, the size of the difference between Eqs.~(\ref{eq:DIPSAE}) and 
(\ref{eq:DIPCIA}) depends on the magnitude of $C_2$ and the relative 
difference between $\langle \epsilon\sigma_g | \op{D} | 1\sigma_u  \rangle$ 
and $\langle 1\sigma_u | \op{D} | 1\sigma_g  \rangle$. Although correlation 
in the final state and thus $C_2$ is expected to be small, it can  
be compensated by the size of the dipole matrix element. Using the 
additional simplification of adopting H$_2^+$ orbitals instead of 
the ones of H$_2$ one finds (with the CI$^{\rm a}$ basis) the maximum
value of the coefficient $C_2$ to be about 0.15, and thus the
maximum contribution of the {\susu} configuration to the
continuum wavefunction to amount to about 2.25\%. While this 
coefficient is very small, the dipole moment 
$\langle 1\sigma_u | \op{D} | 1\sigma_g \rangle$ between two bound orbitals 
of similar spatial extension is on the other hand much larger than the 
one between the continuum orbital $\epsilon\sigma_g$ and the bound 
state $1\sigma_u$. While the former is found to be equal to 
1.1602\,a.\,u.\ at $R=2.3\,a_0$ (in good agreement with the results 
in~\cite{dia:bate51, sfm:lin00c}), the discretized (and thus equally 
normalized) transition dipole matrix element 
$\langle n\sigma_g | \op{D} | 1\sigma_u \rangle$ is found to be 
0.0408\,a.\,u.\ at a final-state energy corresponding to a photon 
energy of about 8\,eV. This explains easily the dominance of the 
second amplitude in Eq.~(\ref{eq:DIPCIA}) and the failure of an 
SAE model. 

Even a very small admixture of states like 
$|1\sigma_u \rangle |1\sigma_u \rangle$ due to correlation can 
substantially modify the one-photon cross section, since the bound-bound 
transition matrix elements are much larger than the bound-continuum 
ones. From Fig.~\ref{fig:H2_R2p3_SPI_CRS} one may further 
conclude that the coefficient $C_2$ changes sign (relative to $C_1$) 
at a final-state energy at about 2 to 3\,eV above the ionization 
threshold, since the cross section is first reduced and than 
enhanced compared to the one obtained with the SAE model.
Clearly, a comparable effect does not occur for the electronic 
ground state, since all matrix elements of the type 
$\langle n\sigma_g ; m \sigma_u | \op{D} 
                              | 1\sigma_g ; 1\sigma_g \rangle$ 
vanish systematically for $n\neq 1$ due to the orthogonality of the 
orbitals; the matrix element of a one-electron operator between two 
Slater determinants differing by more than one orbital is zero. 
%
%
%
%
%
\section{Conclusions}\label{sec:conclusion}
On the basis of results obtained from the solution of full dimensional 
time-dependent Schr\"odinger equation describing the two electrons 
of H$_2$ exposed to an intense laser field the ionization of the 
first electronically excited {\bsig} state is investigated in the 
one-photon regime. A large deviation is found compared to a 
corresponding recent calculation \cite{sfm:bart08} performed within 
a Koopmans' picture based SAE approximation. It is demonstrated that also 
two other SAE approximations (called p-SAE and DFT-SAE) fail similarly, 
although DFT-SAE seemed on the first glance to yield reasonable 
results. However, this turned out to be accidentally caused by 
two partially compensating errors. On the other hand, lowest-order 
perturbation theory was shown to be quite adequate to reproduce the 
full TDSE calculations. This allowed to demonstrate that the failure 
of the SAE approximations is solely due to an improper description of the 
final continuum state, while the initial {\bsig} state is properly 
described. The inclusion of a single doubly-excited configuration into 
the configuration-interaction calculation of the final states explains 
semi-quantitatively the increase of the cross section by about an order 
of magnitude compared to an SAE model. The present work is thus an 
interesting example for a dramatic effect of correlation in the final 
continuum states that emerges for an excited initial electronic state 
but is absent, if the initial state is the electronic ground state 
of H$_2$. The found effect may be observable with synchrotron radiation. 
However, since it is rather difficult to achieve the required target 
density of electronically excited states, experiments with higher 
photon fluxes as they are available from free-electron lasers 
should be advantageous. 

%
%
\section*{Acknowledgments}
The authors acknowledge numerous helpful discussions with Ingo Barth 
and J\"orn Manz and financial support by the {\it Deutsche 
Forschungsgemeinschaft (DFG)} within SFB~450 (C6) as well as within 
Grant No.~Sa936/2, by the COST program CM0702, and by the 
{\it Fonds der Chemischen Industrie}. This research was also 
supported in part by the National Science Foundation under Grant 
No.~PHY05-51164.

%
%


\begin{thebibliography}{10}%
\makeatletter
\providecommand \@ifxundefined [1]{%
 \ifx #1\undefined \expandafter \@firstoftwo
 \else \expandafter \@secondoftwo
\fi
}%
\providecommand \@ifnum [1]{%
 \ifnum #1\expandafter \@firstoftwo
 \else \expandafter \@secondoftwo
\fi
}%
\providecommand \enquote [1]{``#1''}%
\providecommand \bibnamefont  [1]{#1}%
\providecommand \bibfnamefont [1]{#1}%
\providecommand \citenamefont [1]{#1}%
\providecommand\href[0]{\@sanitize\@href}%
\providecommand\@href[1]{\endgroup\@@startlink{#1}\endgroup\@@href}%
\providecommand\@@href[1]{#1\@@endlink}%
\providecommand \@sanitize [0]{\begingroup\catcode`\&12\catcode`\#12\relax}%
\@ifxundefined \pdfoutput {\@firstoftwo}{%
 \@ifnum{\z@=\pdfoutput}{\@firstoftwo}{\@secondoftwo}%
}{%
 \providecommand\@@startlink[1]{\leavevmode}%
 \providecommand\@@endlink[0]{}%
}{%
 \providecommand\@@startlink[1]{%
  \leavevmode
  \pdfstartlink
   attr{/Border[0 0 1 ]/H/I/C[0 1 1]}%
   user{/Subtype/Link/A<</Type/Action/S/URI/URI(#1)>>}%
  \relax
 }%
 \providecommand\@@endlink[0]{\pdfendlink}%
}%
\providecommand \url  [0]{\begingroup\@sanitize \@url }%
\providecommand \@url [1]{\endgroup\@href {#1}{\urlprefix}}%
\providecommand \urlprefix [0]{URL }%
\providecommand \Eprint[0]{\href }%
\@ifxundefined \urlstyle {%
  \providecommand \doi [1]{doi:\discretionary{}{}{}#1}%
}{%
  \providecommand \doi [0]{doi:\discretionary{}{}{}\begingroup
  \urlstyle{rm}\Url }%
}%
\providecommand \doibase [0]{http://dx.doi.org/}%
\providecommand \Doi[1]{\href{\doibase#1}}%
\providecommand \bibAnnote [3]{%
  \BibitemShut{#1}%
  \begin{quotation}\noindent
    \textsc{Key:}\ #2\\\textsc{Annotation:}\ #3%
  \end{quotation}%
}%
\providecommand \bibAnnoteFile [2]{%
  \IfFileExists{#2}{\bibAnnote {#1} {#2} {\input{#2}}}{}%
}%
\providecommand \typeout [0]{\immediate \write \m@ne }%
\providecommand \selectlanguage [0]{\@gobble}%
\providecommand \bibinfo [0]{\@secondoftwo}%
\providecommand \bibfield [0]{\@secondoftwo}%
\providecommand \translation [1]{[#1]}%
\providecommand \BibitemOpen[0]{}%
\providecommand \bibitemStop [0]{}%
\providecommand \bibitemNoStop [0]{.\EOS\space}%
\providecommand \EOS [0]{\spacefactor3000\relax}%
\providecommand \BibitemShut [1]{\csname bibitem#1\endcsname}%
\bibitem{gen:made71}%
  \BibitemOpen
  \bibfield{author}{%
  \bibinfo {author} {\bibfnamefont{J.~M.~J.}\ \bibnamefont{Madey}},\ }%
  \bibfield{journal}{%
  {\bibinfo {journal} {Journal of Applied Physics}}\ }%
  \textbf{\bibinfo {volume} {42}},\ \bibinfo {pages} {1906} (\bibinfo {year}
  {1971})%
  \bibAnnoteFile{NoStop}{gen:made71}%
\bibitem{sfa:rude08}%
  \BibitemOpen
  \bibfield{author}{%
  \bibinfo {author} {\bibfnamefont{A.}~\bibnamefont{Rudenko}}, \bibinfo
  {author} {\bibfnamefont{L.}~\bibnamefont{Foucar}}, \bibinfo {author}
  {\bibfnamefont{M.}~\bibnamefont{Kurka}}, \bibinfo {author}
  {\bibfnamefont{T.}~\bibnamefont{Ergler}}, \bibinfo {author}
  {\bibfnamefont{K.~U.}\ \bibnamefont{K\"{u}hnel}}, \bibinfo {author}
  {\bibfnamefont{Y.~H.}\ \bibnamefont{Jiang}}, \bibinfo {author}
  {\bibfnamefont{A.}~\bibnamefont{Voitkiv}}, \bibinfo {author}
  {\bibfnamefont{B.}~\bibnamefont{Najjari}}, \bibinfo {author}
  {\bibfnamefont{A.}~\bibnamefont{Kheifets}}, \bibinfo {author}
  {\bibfnamefont{S.}~\bibnamefont{L\"{u}demann}}, \bibinfo {author}
  {\bibfnamefont{T.}~\bibnamefont{Havermeier}}, \bibinfo {author}
  {\bibfnamefont{M.}~\bibnamefont{Smolarski}}, \bibinfo {author}
  {\bibfnamefont{S.}~\bibnamefont{Sch\"{o}ssler}}, \bibinfo {author}
  {\bibfnamefont{K.}~\bibnamefont{Cole}}, \bibinfo {author}
  {\bibfnamefont{M.}~\bibnamefont{Sch\"{o}ffler}}, \bibinfo {author}
  {\bibfnamefont{R.}~\bibnamefont{D\"{o}rner}}, \bibinfo {author}
  {\bibfnamefont{S.}~\bibnamefont{D\"{u}sterer}}, \bibinfo {author}
  {\bibfnamefont{W.}~\bibnamefont{Li}}, \bibinfo {author}
  {\bibfnamefont{B.}~\bibnamefont{Keitel}}, \bibinfo {author}
  {\bibfnamefont{R.}~\bibnamefont{Treusch}}, \bibinfo {author}
  {\bibfnamefont{M.}~\bibnamefont{Gensch}}, \bibinfo {author}
  {\bibfnamefont{C.~D.}\ \bibnamefont{Schr\"{o}ter}}, \bibinfo {author}
  {\bibfnamefont{R.}~\bibnamefont{Moshammer}},\ and\ \bibinfo {author}
  {\bibfnamefont{J.}~\bibnamefont{Ullrich}},\ }%
  \bibfield{journal}{%
  \bibinfo {journal} {Phys.\,Rev.\,Lett.}\ }%
  \textbf{\bibinfo {volume} {101}},\ \bibinfo {pages} {073003} (\bibinfo {year}
  {2008})%
  \bibAnnoteFile{NoStop}{sfa:rude08}%
\bibitem{sfa:rich09}%
  \BibitemOpen
  \bibfield{author}{%
  \bibinfo {author} {\bibfnamefont{M.}~\bibnamefont{Richter}}, \bibinfo
  {author} {\bibfnamefont{M.~Y.}\ \bibnamefont{Amusia}}, \bibinfo {author}
  {\bibfnamefont{S.~V.}\ \bibnamefont{Bobashev}}, \bibinfo {author}
  {\bibfnamefont{T.}~\bibnamefont{Feigl}}, \bibinfo {author}
  {\bibfnamefont{P.~N.}\ \bibnamefont{Jurani\'{c}}}, \bibinfo {author}
  {\bibfnamefont{M.}~\bibnamefont{Martins}}, \bibinfo {author}
  {\bibfnamefont{A.~A.}\ \bibnamefont{Sorokin}},\ and\ \bibinfo {author}
  {\bibfnamefont{K.}~\bibnamefont{Tiedtke}},\ }%
  \bibfield{journal}{%
  \bibinfo {journal} {Phys.\,Rev.\,Lett.}\ }%
  \textbf{\bibinfo {volume} {102}},\ \bibinfo {pages} {163002} (\bibinfo {year}
  {2009})%
  \bibAnnoteFile{NoStop}{sfa:rich09}%
\bibitem{sfm:jian09}%
  \BibitemOpen
  \bibfield{author}{%
  \bibinfo {author} {\bibfnamefont{Y.~H.}\ \bibnamefont{Jiang}}, \bibinfo
  {author} {\bibfnamefont{A.}~\bibnamefont{Rudenko}}, \bibinfo {author}
  {\bibfnamefont{M.}~\bibnamefont{Kurka}}, \bibinfo {author}
  {\bibfnamefont{K.~U.}\ \bibnamefont{K{\"u}hnel}}, \bibinfo {author}
  {\bibfnamefont{T.}~\bibnamefont{Ergler}}, \bibinfo {author}
  {\bibfnamefont{L.}~\bibnamefont{Foucar}}, \bibinfo {author}
  {\bibfnamefont{M.}~\bibnamefont{Sch{\"o}ffler}}, \bibinfo {author}
  {\bibfnamefont{S.}~\bibnamefont{Sch{\"o}ssler}}, \bibinfo {author}
  {\bibfnamefont{T.}~\bibnamefont{Havermeier}}, \bibinfo {author}
  {\bibfnamefont{M.}~\bibnamefont{Smolarski}}, \bibinfo {author}
  {\bibfnamefont{K.}~\bibnamefont{Cole}}, \bibinfo {author}
  {\bibfnamefont{R.}~\bibnamefont{D{\"o}rner}}, \bibinfo {author}
  {\bibfnamefont{S.}~\bibnamefont{D{\"u}sterer}}, \bibinfo {author}
  {\bibfnamefont{R.}~\bibnamefont{Treusch}}, \bibinfo {author}
  {\bibfnamefont{M.}~\bibnamefont{Gensch}}, \bibinfo {author}
  {\bibfnamefont{C.~D.}\ \bibnamefont{Schr{\"o}ter}}, \bibinfo {author}
  {\bibfnamefont{R.}~\bibnamefont{Moshammer}},\ and\ \bibinfo {author}
  {\bibfnamefont{J.}~\bibnamefont{Ullrich}},\ }%
  \bibfield{journal}{%
  \bibinfo {journal} {Phys.\,Rev.\,Lett.}\ }%
  \textbf{\bibinfo {volume} {102}},\ \bibinfo {pages} {123002} (\bibinfo {year}
  {2009})%
  \bibAnnoteFile{NoStop}{sfm:jian09}%
\bibitem{sfa:zhu09}%
  \BibitemOpen
  \bibfield{author}{%
  \bibinfo {author} {\bibfnamefont{G.}~\bibnamefont{Zhu}}, \bibinfo {author}
  {\bibfnamefont{M.}~\bibnamefont{Schuricke}}, \bibinfo {author}
  {\bibfnamefont{J.}~\bibnamefont{Steinmann}}, \bibinfo {author}
  {\bibfnamefont{J.}~\bibnamefont{Albrecht}}, \bibinfo {author}
  {\bibfnamefont{J.}~\bibnamefont{Ullrich}}, \bibinfo {author}
  {\bibfnamefont{I.}~\bibnamefont{Ben-Itzhak}}, \bibinfo {author}
  {\bibfnamefont{T.~J.~M.}\ \bibnamefont{Zouros}}, \bibinfo {author}
  {\bibfnamefont{J.}~\bibnamefont{Colgan}}, \bibinfo {author}
  {\bibfnamefont{M.~S.}\ \bibnamefont{Pindzola}},\ and\ \bibinfo {author}
  {\bibfnamefont{A.}~\bibnamefont{Dorn}},\ }%
  \bibfield{journal}{%
  \bibinfo {journal} {Phys.\,Rev.\,Lett.}\ }%
  \textbf{\bibinfo {volume} {103}},\ \bibinfo {pages} {103008} (\bibinfo {year}
  {2009})%
  \bibAnnoteFile{NoStop}{sfa:zhu09}%
\bibitem{sfa:keld65}%
  \BibitemOpen
  \bibfield{author}{%
  \bibinfo {author} {\bibfnamefont{L.~V.}\ \bibnamefont{Keldysh}},\ }%
  \bibfield{journal}{%
  \bibinfo {journal} {Sov.\,Phys.\ JETP}\ }%
  \textbf{\bibinfo {volume} {20}},\ \bibinfo {pages} {1307} (\bibinfo {year}
  {1965})%
  \bibAnnoteFile{NoStop}{sfa:keld65}%
\bibitem{sfm:apal01}%
  \BibitemOpen
  \bibfield{author}{%
  \bibinfo {author} {\bibfnamefont{A.}~\bibnamefont{Apalategui}}, \bibinfo
  {author} {\bibfnamefont{A.}~\bibnamefont{Saenz}},\ and\ \bibinfo {author}
  {\bibfnamefont{P.}~\bibnamefont{Lambropoulos}},\ }%
  \bibfield{journal}{%
  \bibinfo {journal} {Phys.\,Rev.\,Lett.}\ }%
  \textbf{\bibinfo {volume} {86}},\ \bibinfo {pages} {5454} (\bibinfo {year}
  {2001})%
  \bibAnnoteFile{NoStop}{sfm:apal01}%
\bibitem{sfm:apal02}%
  \BibitemOpen
  \bibfield{author}{%
  \bibinfo {author} {\bibfnamefont{A.}~\bibnamefont{Apalategui}}\ and\ \bibinfo
  {author} {\bibfnamefont{A.}~\bibnamefont{Saenz}},\ }%
  \bibfield{journal}{%
  \bibinfo {journal} {J.\,Phys.\ B}\ }%
  \textbf{\bibinfo {volume} {35}},\ \bibinfo {pages} {1909} (\bibinfo {year}
  {2002})%
  \bibAnnoteFile{NoStop}{sfm:apal02}%
\bibitem{sfm:kawa00}%
  \BibitemOpen
  \bibfield{author}{%
  \bibinfo {author} {\bibfnamefont{I.}~\bibnamefont{Kawata}}, \bibinfo {author}
  {\bibfnamefont{H.}~\bibnamefont{Kono}}, \bibinfo {author}
  {\bibfnamefont{Y.}~\bibnamefont{Fujimura}},\ and\ \bibinfo {author}
  {\bibfnamefont{A.~D.}\ \bibnamefont{Bandrauk}},\ }%
  \bibfield{journal}{%
  \bibinfo {journal} {Phys.\,Rev.\ A}\ }%
  \textbf{\bibinfo {volume} {62}},\ \bibinfo {pages} {031401(R)} (\bibinfo
  {year} {2000})%
  \bibAnnoteFile{NoStop}{sfm:kawa00}%
\bibitem{sfm:awas05}%
  \BibitemOpen
  \bibfield{author}{%
  \bibinfo {author} {\bibfnamefont{M.}~\bibnamefont{Awasthi}}, \bibinfo
  {author} {\bibfnamefont{Y.~V.}\ \bibnamefont{Vanne}},\ and\ \bibinfo {author}
  {\bibfnamefont{A.}~\bibnamefont{Saenz}},\ }%
  \bibfield{journal}{%
  \bibinfo {journal} {J.\,Phys.\ B}\ }%
  \textbf{\bibinfo {volume} {38}},\ \bibinfo {pages} {3973} (\bibinfo {year}
  {2005})%
  \bibAnnoteFile{NoStop}{sfm:awas05}%
\bibitem{sfm:awas06}%
  \BibitemOpen
  \bibfield{author}{%
  \bibinfo {author} {\bibfnamefont{M.}~\bibnamefont{Awasthi}}\ and\ \bibinfo
  {author} {\bibfnamefont{A.}~\bibnamefont{Saenz}},\ }%
  \bibfield{journal}{%
  \bibinfo {journal} {J.\,Phys.\ B}\ }%
  \textbf{\bibinfo {volume} {39}},\ \bibinfo {pages} {S\,389} (\bibinfo {year}
  {2006})%
  \bibAnnoteFile{NoStop}{sfm:awas06}%
\bibitem{sfm:pala06}%
  \BibitemOpen
  \bibfield{author}{%
  \bibinfo {author} {\bibfnamefont{A.}~\bibnamefont{Palacios}}, \bibinfo
  {author} {\bibfnamefont{H.}~\bibnamefont{Bachau}},\ and\ \bibinfo {author}
  {\bibfnamefont{F.}~\bibnamefont{Mart{\'i}n}},\ }%
  \bibfield{journal}{%
  \bibinfo {journal} {Phys.\,Rev.\,Lett.}\ }%
  \textbf{\bibinfo {volume} {96}},\ \bibinfo {pages} {143001} (\bibinfo {year}
  {2006})%
  \bibAnnoteFile{NoStop}{sfm:pala06}%
\bibitem{sfm:vann08}%
  \BibitemOpen
  \bibfield{author}{%
  \bibinfo {author} {\bibfnamefont{Y.~V.}\ \bibnamefont{Vanne}}\ and\ \bibinfo
  {author} {\bibfnamefont{A.}~\bibnamefont{Saenz}},\ }%
  \bibfield{journal}{%
  \bibinfo {journal} {J.\,Mod.\,Opt.}\ }%
  \textbf{\bibinfo {volume} {55}},\ \bibinfo {pages} {2665} (\bibinfo {year}
  {2008})%
  \bibAnnoteFile{NoStop}{sfm:vann08}%
\bibitem{sfm:vann09}%
  \BibitemOpen
  \bibfield{author}{%
  \bibinfo {author} {\bibfnamefont{Y.~V.}\ \bibnamefont{Vanne}}\ and\ \bibinfo
  {author} {\bibfnamefont{A.}~\bibnamefont{Saenz}},\ }%
  \bibfield{journal}{%
  \bibinfo {journal} {Phys.\,Rev.\ A}\ }%
  \textbf{\bibinfo {volume} {80}},\ \bibinfo {pages} {053422} (\bibinfo {year}
  {2009})%
  \bibAnnoteFile{NoStop}{sfm:vann09}%
\bibitem{sfa:debo92}%
  \BibitemOpen
  \bibfield{author}{%
  \bibinfo {author} {\bibfnamefont{M.~P.}\ \bibnamefont{de~Boer}}\ and\
  \bibinfo {author} {\bibfnamefont{H.~G.}\ \bibnamefont{Muller}},\ }%
  \bibfield{journal}{%
  \bibinfo {journal} {Phys.\,Rev.\,Lett.}\ }%
  \textbf{\bibinfo {volume} {68}},\ \bibinfo {pages} {2747} (\bibinfo {year}
  {1992})%
  \bibAnnoteFile{NoStop}{sfa:debo92}%
\bibitem{sfa:jone93}%
  \BibitemOpen
  \bibfield{author}{%
  \bibinfo {author} {\bibfnamefont{R.~R.}\ \bibnamefont{Jones}}, \bibinfo
  {author} {\bibfnamefont{D.~W.}\ \bibnamefont{Schumacher}},\ and\ \bibinfo
  {author} {\bibfnamefont{P.~H.}\ \bibnamefont{Bucksbaum}},\ }%
  \bibfield{journal}{%
  \bibinfo {journal} {Phys.\,Rev.\ A}\ }%
  \textbf{\bibinfo {volume} {47}},\ \bibinfo {pages} {R49} (\bibinfo {year}
  {1993})%
  \bibAnnoteFile{NoStop}{sfa:jone93}%
\bibitem{sfa:hert97}%
  \BibitemOpen
  \bibfield{author}{%
  \bibinfo {author} {\bibfnamefont{M.~P.}\ \bibnamefont{Hertlein}}, \bibinfo
  {author} {\bibfnamefont{P.~H.}\ \bibnamefont{Bucksbaum}},\ and\ \bibinfo
  {author} {\bibfnamefont{H.~G.}\ \bibnamefont{Muller}},\ }%
  \bibfield{journal}{%
  \bibinfo {journal} {J.\,Phys.\ B}\ }%
  \textbf{\bibinfo {volume} {30}},\ \bibinfo {pages} {L197} (\bibinfo {year}
  {1997})%
  \bibAnnoteFile{NoStop}{sfa:hert97}%
\bibitem{sfa:mull99}%
  \BibitemOpen
  \bibfield{author}{%
  \bibinfo {author} {\bibfnamefont{H.~G.}\ \bibnamefont{Muller}},\ }%
  \bibfield{journal}{%
  \bibinfo {journal} {Laser Phys.}\ }%
  \textbf{\bibinfo {volume} {9}},\ \bibinfo {pages} {138} (\bibinfo {year}
  {1999})%
  \bibAnnoteFile{NoStop}{sfa:mull99}%
\bibitem{sfa:nubb08}%
  \BibitemOpen
  \bibfield{author}{%
  \bibinfo {author} {\bibfnamefont{T.}~\bibnamefont{Nubbemeyer}}, \bibinfo
  {author} {\bibfnamefont{K.}~\bibnamefont{Gorling}}, \bibinfo {author}
  {\bibfnamefont{A.}~\bibnamefont{Saenz}}, \bibinfo {author}
  {\bibfnamefont{U.}~\bibnamefont{Eichmann}},\ and\ \bibinfo {author}
  {\bibfnamefont{W.}~\bibnamefont{Sandner}},\ }%
  \bibfield{journal}{%
  \bibinfo {journal} {Phys.\,Rev.\,Lett.}\ }%
  \textbf{\bibinfo {volume} {101}},\ \bibinfo {pages} {233001} (\bibinfo {year}
  {2008})%
  \bibAnnoteFile{NoStop}{sfa:nubb08}%
\bibitem{sfm:awas08}%
  \BibitemOpen
  \bibfield{author}{%
  \bibinfo {author} {\bibfnamefont{M.}~\bibnamefont{Awasthi}}, \bibinfo
  {author} {\bibfnamefont{Y.~V.}\ \bibnamefont{Vanne}}, \bibinfo {author}
  {\bibfnamefont{A.}~\bibnamefont{Saenz}}, \bibinfo {author}
  {\bibfnamefont{A.}~\bibnamefont{Castro}},\ and\ \bibinfo {author}
  {\bibfnamefont{P.}~\bibnamefont{Decleva}},\ }%
  \bibfield{journal}{%
  \bibinfo {journal} {Phys.\,Rev.\ A}\ }%
  \textbf{\bibinfo {volume} {77}},\ \bibinfo {pages} {063403} (\bibinfo {year}
  {2008})%
  \bibAnnoteFile{NoStop}{sfm:awas08}%
\bibitem{sfm:mans09}%
  \BibitemOpen
  \bibfield{author}{%
  \bibinfo {author} {\bibfnamefont{B.}~\bibnamefont{Manschwetus}}, \bibinfo
  {author} {\bibfnamefont{T.}~\bibnamefont{Nubbemeyer}}, \bibinfo {author}
  {\bibfnamefont{K.}~\bibnamefont{Gorling}}, \bibinfo {author}
  {\bibfnamefont{G.}~\bibnamefont{Steinmeyer}}, \bibinfo {author}
  {\bibfnamefont{U.}~\bibnamefont{Eichmann}}, \bibinfo {author}
  {\bibfnamefont{H.}~\bibnamefont{Rottke}},\ and\ \bibinfo {author}
  {\bibfnamefont{W.}~\bibnamefont{Sandner}},\ }%
  \bibfield{journal}{%
  \bibinfo {journal} {Phys.\,Rev.\,Lett.}\ }%
  \textbf{\bibinfo {volume} {102}},\ \bibinfo {pages} {113002} (\bibinfo {year}
  {2009})%
  \bibAnnoteFile{NoStop}{sfm:mans09}%
\bibitem{dia:pede07}%
  \BibitemOpen
  \bibfield{author}{%
  \bibinfo {author} {\bibfnamefont{H.~B.}\ \bibnamefont{Pedersen}}, \bibinfo
  {author} {\bibfnamefont{S.}~\bibnamefont{Altevogt}}, \bibinfo {author}
  {\bibfnamefont{B.}~\bibnamefont{Jordon-Thaden}}, \bibinfo {author}
  {\bibfnamefont{O.}~\bibnamefont{Heber}}, \bibinfo {author}
  {\bibfnamefont{M.~L.}\ \bibnamefont{Rappaport}}, \bibinfo {author}
  {\bibfnamefont{D.}~\bibnamefont{Schwalm}}, \bibinfo {author}
  {\bibfnamefont{J.}~\bibnamefont{Ullrich}}, \bibinfo {author}
  {\bibfnamefont{D.}~\bibnamefont{Zajfman}}, \bibinfo {author}
  {\bibfnamefont{R.}~\bibnamefont{Treusch}}, \bibinfo {author}
  {\bibfnamefont{N.}~\bibnamefont{Guerassimova}}, \bibinfo {author}
  {\bibfnamefont{M.}~\bibnamefont{Martins}}, \bibinfo {author}
  {\bibfnamefont{J.-T.}\ \bibnamefont{Hoeft}}, \bibinfo {author}
  {\bibfnamefont{M.}~\bibnamefont{Wellh{\'o}fer}},\ and\ \bibinfo {author}
  {\bibfnamefont{A.}~\bibnamefont{Wolf}},\ }%
  \bibfield{journal}{%
  \bibinfo {journal} {Phys.\,Rev.\,Lett.}\ }%
  \textbf{\bibinfo {volume} {98}},\ \bibinfo {pages} {223202} (\bibinfo {year}
  {2007})%
  \bibAnnoteFile{NoStop}{dia:pede07}%
\bibitem{dia:dumi09}%
  \BibitemOpen
  \bibfield{author}{%
  \bibinfo {author} {\bibfnamefont{I.}~\bibnamefont{Dumitriu}}\ and\ \bibinfo
  {author} {\bibfnamefont{A.}~\bibnamefont{Saenz}},\ }%
  \bibfield{journal}{%
  \bibinfo {journal} {J.\,Phys.\ B}\ }%
  \textbf{\bibinfo {volume} {42}},\ \bibinfo {pages} {165101} (\bibinfo {year}
  {2009})%
  \bibAnnoteFile{NoStop}{dia:dumi09}%
\bibitem{sfm:saen02c}%
  \BibitemOpen
  \bibfield{author}{%
  \bibinfo {author} {\bibfnamefont{A.}~\bibnamefont{Saenz}},\ }%
  \bibfield{journal}{%
  \bibinfo {journal} {J.\,Phys.\ B}\ }%
  \textbf{\bibinfo {volume} {35}},\ \bibinfo {pages} {4829} (\bibinfo {year}
  {2002})%
  \bibAnnoteFile{NoStop}{sfm:saen02c}%
\bibitem{sfm:bart08}%
  \BibitemOpen
  \bibfield{author}{%
  \bibinfo {author} {\bibfnamefont{I.}~\bibnamefont{Barth}}, \bibinfo {author}
  {\bibfnamefont{J.}~\bibnamefont{Manz}},\ and\ \bibinfo {author}
  {\bibfnamefont{G.~K.}\ \bibnamefont{Paramonov}},\ }%
  \bibfield{journal}{%
  \bibinfo {journal} {Mol.\,Phys.}\ }%
  \textbf{\bibinfo {volume} {106}},\ \bibinfo {pages} {467} (\bibinfo {year}
  {2008})%
  \bibAnnoteFile{NoStop}{sfm:bart08}%
\bibitem{gen:andr00}%
  \BibitemOpen
  \bibfield{author}{%
  \bibinfo {author} {\bibfnamefont{J.}~\bibnamefont{Andruszkow}}, \bibinfo
  {author} {\bibfnamefont{B.}~\bibnamefont{Aune}}, \bibinfo {author}
  {\bibfnamefont{V.}~\bibnamefont{Ayvazyan}}, \bibinfo {author}
  {\bibfnamefont{N.}~\bibnamefont{Baboi}}, \bibinfo {author}
  {\bibfnamefont{R.}~\bibnamefont{Bakker}}, \bibinfo {author}
  {\bibfnamefont{V.}~\bibnamefont{Balakin}}, \bibinfo {author}
  {\bibfnamefont{D.}~\bibnamefont{Barni}}, \bibinfo {author}
  {\bibfnamefont{A.}~\bibnamefont{Bazhan}}, \bibinfo {author}
  {\bibfnamefont{M.}~\bibnamefont{Bernard}}, \bibinfo {author}
  {\bibfnamefont{A.}~\bibnamefont{Bosotti}}, \bibinfo {author}
  {\bibfnamefont{J.~C.}\ \bibnamefont{Bourdon}}, \bibinfo {author}
  {\bibfnamefont{W.}~\bibnamefont{Brefeld}}, \bibinfo {author}
  {\bibfnamefont{R.}~\bibnamefont{Brinkmann}}, \bibinfo {author}
  {\bibfnamefont{S.}~\bibnamefont{Buhler}}, \bibinfo {author}
  {\bibfnamefont{J.-P.}\ \bibnamefont{Carneiro}}, \bibinfo {author}
  {\bibfnamefont{M.}~\bibnamefont{Castellano}}, \bibinfo {author}
  {\bibfnamefont{P.}~\bibnamefont{Castro}}, \bibinfo {author}
  {\bibfnamefont{L.}~\bibnamefont{Catani}}, \bibinfo {author}
  {\bibfnamefont{S.}~\bibnamefont{Chel}}, \bibinfo {author}
  {\bibfnamefont{Y.}~\bibnamefont{Cho}}, \bibinfo {author}
  {\bibfnamefont{S.}~\bibnamefont{Choroba}}, \bibinfo {author}
  {\bibfnamefont{E.~R.}\ \bibnamefont{Colby}}, \bibinfo {author}
  {\bibfnamefont{W.}~\bibnamefont{Decking}}, \bibinfo {author}
  {\bibfnamefont{P.}~\bibnamefont{Den~Hartog}}, \bibinfo {author}
  {\bibfnamefont{M.}~\bibnamefont{Desmons}}, \bibinfo {author}
  {\bibfnamefont{M.}~\bibnamefont{Dohlus}},\ and\ \bibinfo {author}
  {\bibfnamefont{D.}~\bibnamefont{Edwards}},\ }%
  \bibfield{journal}{%
  {\bibinfo {journal} {Phys.\,Rev.\,Lett.}}\
  }%
  \textbf{\bibinfo {volume} {85}},\ \bibinfo {pages} {3825} (\bibinfo {year}
  {2000})%
  \bibAnnoteFile{NoStop}{gen:andr00}%
\bibitem{gen:yu03}%
  \BibitemOpen
  \bibfield{author}{%
  \bibinfo {author} {\bibfnamefont{L.~H.}\ \bibnamefont{Yu}}, \bibinfo {author}
  {\bibfnamefont{L.}~\bibnamefont{DiMauro}}, \bibinfo {author}
  {\bibfnamefont{A.}~\bibnamefont{Doyuran}}, \bibinfo {author}
  {\bibfnamefont{W.~S.}\ \bibnamefont{Graves}}, \bibinfo {author}
  {\bibfnamefont{E.~D.}\ \bibnamefont{Johnson}}, \bibinfo {author}
  {\bibfnamefont{R.}~\bibnamefont{Heese}}, \bibinfo {author}
  {\bibfnamefont{S.}~\bibnamefont{Krinsky}}, \bibinfo {author}
  {\bibfnamefont{H.}~\bibnamefont{Loos}}, \bibinfo {author}
  {\bibfnamefont{J.~B.}\ \bibnamefont{Murphy}}, \bibinfo {author}
  {\bibfnamefont{G.}~\bibnamefont{Rakowsky}}, \bibinfo {author}
  {\bibfnamefont{J.}~\bibnamefont{Rose}}, \bibinfo {author}
  {\bibfnamefont{T.}~\bibnamefont{Shaftan}}, \bibinfo {author}
  {\bibfnamefont{B.}~\bibnamefont{Sheehy}}, \bibinfo {author}
  {\bibfnamefont{J.}~\bibnamefont{Skaritka}}, \bibinfo {author}
  {\bibfnamefont{X.~J.}\ \bibnamefont{Wang}},\ and\ \bibinfo {author}
  {\bibfnamefont{Z.}~\bibnamefont{Wu}},\ }%
  \bibfield{journal}{%
  \bibinfo {journal} {Phys.\,Rev.\,Lett.}\ }%
  \textbf{\bibinfo {volume} {91}},\ \bibinfo {pages} {074801} (\bibinfo {year}
  {2003})%
  \bibAnnoteFile{NoStop}{gen:yu03}%
\bibitem{sfa:lamb98a}%
  \BibitemOpen
  \bibfield{author}{%
  \bibinfo {author} {\bibfnamefont{P.}~\bibnamefont{Lambropoulos}}, \bibinfo
  {author} {\bibfnamefont{P.}~\bibnamefont{Maragakis}},\ and\ \bibinfo {author}
  {\bibfnamefont{J.}~\bibnamefont{Zhang}},\ }%
  \bibfield{journal}{%
  \bibinfo {journal} {Phys.\,Rep.}\ }%
  \textbf{\bibinfo {volume} {305}},\ \bibinfo {pages} {203} (\bibinfo {year}
  {1998})%
  \bibAnnoteFile{NoStop}{sfa:lamb98a}%
\bibitem{dia:kolo65}%
  \BibitemOpen
  \bibfield{author}{%
  \bibinfo {author} {\bibfnamefont{W.}~\bibnamefont{Kolos}}\ and\ \bibinfo
  {author} {\bibfnamefont{L.}~\bibnamefont{Wolniewicz}},\ }%
  \bibfield{journal}{%
  \bibinfo {journal} {J.\,Chem.\,Phys.}\ }%
  \textbf{\bibinfo {volume} {43}},\ \bibinfo {pages} {2429} (\bibinfo {year}
  {1965})%
  \bibAnnoteFile{NoStop}{dia:kolo65}%
\bibitem{dia:woln88}%
  \BibitemOpen
  \bibfield{author}{%
  \bibinfo {author} {\bibfnamefont{L.}~\bibnamefont{Wolniewicz}}\ and\ \bibinfo
  {author} {\bibfnamefont{K.}~\bibnamefont{Dressler}},\ }%
  \bibfield{journal}{%
  \bibinfo {journal} {J.\,Chem.\,Phys.}\ }%
  \textbf{\bibinfo {volume} {88}},\ \bibinfo {pages} {3861} (\bibinfo {year}
  {1988})%
  \bibAnnoteFile{NoStop}{dia:woln88}%
\bibitem{dia:vann04}%
  \BibitemOpen
  \bibfield{author}{%
  \bibinfo {author} {\bibfnamefont{Y.~V.}\ \bibnamefont{Vanne}}\ and\ \bibinfo
  {author} {\bibfnamefont{A.}~\bibnamefont{Saenz}},\ }%
  \bibfield{journal}{%
  \bibinfo {journal} {J.\,Phys.\ B}\ }%
  \textbf{\bibinfo {volume} {37}},\ \bibinfo {pages} {4101} (\bibinfo {year}
  {2004})%
  \bibAnnoteFile{NoStop}{dia:vann04}%
\bibitem{dia:vann06}%
  \BibitemOpen
  \bibfield{author}{%
  \bibinfo {author} {\bibfnamefont{Y.~V.}\ \bibnamefont{Vanne}}, \bibinfo
  {author} {\bibfnamefont{A.}~\bibnamefont{Saenz}}, \bibinfo {author}
  {\bibfnamefont{A.}~\bibnamefont{Dalgarno}}, \bibinfo {author}
  {\bibfnamefont{R.~C.}\ \bibnamefont{Forrey}}, \bibinfo {author}
  {\bibfnamefont{P.}~\bibnamefont{Froelich}},\ and\ \bibinfo {author}
  {\bibfnamefont{S.}~\bibnamefont{Jonsell}},\ }%
  \bibfield{journal}{%
  \bibinfo {journal} {Phys.\,Rev.\ A}\ }%
  \textbf{\bibinfo {volume} {73}},\ \bibinfo {pages} {062706} (\bibinfo {year}
  {2006})%
  \bibAnnoteFile{NoStop}{dia:vann06}%
\bibitem{sfa:lamb00}%
  \BibitemOpen
  \bibfield{author}{%
  \bibinfo {author} {\bibfnamefont{P.}~\bibnamefont{Lambropoulos}}, \bibinfo
  {author} {\bibfnamefont{M.~A.}\ \bibnamefont{Kornberg}}, \bibinfo {author}
  {\bibfnamefont{L.~A.~A.}\ \bibnamefont{Nikolopoulos}},\ and\ \bibinfo
  {author} {\bibfnamefont{A.}~\bibnamefont{Saenz}},\ }%
  in\ \emph{\bibinfo {booktitle} {Multiphoton Processes}},\ \bibinfo {editor}
  {edited by\ \bibinfo {editor} {\bibfnamefont{L.~F.}\
  \bibnamefont{{DiMauro}}}, \bibinfo {editor} {\bibfnamefont{R.~R.}\
  \bibnamefont{Freeman}},\ and\ \bibinfo {editor} {\bibfnamefont{K.~C.}\
  \bibnamefont{Kulander}}}\ (\bibinfo {publisher} {Melville},\ \bibinfo
  {address} {New York},\ \bibinfo {year} {2000})\ p.\ \bibinfo {pages} {231}%
  \bibAnnoteFile{NoStop}{sfa:lamb00}%
\bibitem{dia:shar71}%
  \BibitemOpen
  \bibfield{author}{%
  \bibinfo {author} {\bibfnamefont{T.~E.}\ \bibnamefont{Sharp}},\ }%
  \bibfield{journal}{%
  \bibinfo {journal} {Atomic Data}\ }%
  \textbf{\bibinfo {volume} {2}},\ \bibinfo {pages} {119} (\bibinfo {year}
  {1971})%
  \bibAnnoteFile{NoStop}{dia:shar71}%
\bibitem{sfm:saen00a}%
  \BibitemOpen
  \bibfield{author}{%
  \bibinfo {author} {\bibfnamefont{A.}~\bibnamefont{Saenz}},\ }%
  \bibfield{journal}{%
  \bibinfo {journal} {Phys.\,Rev.\ A}\ }%
  \textbf{\bibinfo {volume} {61}},\ \bibinfo {pages} {051402(R)} (\bibinfo
  {year} {2000})%
  \bibAnnoteFile{NoStop}{sfm:saen00a}%
\bibitem{sfm:saen00b}%
  \BibitemOpen
  \bibfield{author}{%
  \bibinfo {author} {\bibfnamefont{A.}~\bibnamefont{Saenz}},\ }%
  \bibfield{journal}{%
  \bibinfo {journal} {J.\,Phys.\ B}\ }%
  \textbf{\bibinfo {volume} {33}},\ \bibinfo {pages} {3519} (\bibinfo {year}
  {2000})%
  \bibAnnoteFile{NoStop}{sfm:saen00b}%
\bibitem{sfm:haru00}%
  \BibitemOpen
  \bibfield{author}{%
  \bibinfo {author} {\bibfnamefont{K.}~\bibnamefont{Harumiya}}, \bibinfo
  {author} {\bibfnamefont{I.}~\bibnamefont{Kawata}}, \bibinfo {author}
  {\bibfnamefont{H.}~\bibnamefont{Kono}},\ and\ \bibinfo {author}
  {\bibfnamefont{Y.}~\bibnamefont{Fujimura}},\ }%
  \bibfield{journal}{%
  \bibinfo {journal} {J.\,Chem.\,Phys.}\ }%
  \textbf{\bibinfo {volume} {113}},\ \bibinfo {pages} {8953} (\bibinfo {year}
  {2000})%
  \bibAnnoteFile{NoStop}{sfm:haru00}%
\bibitem{sfm:saen02a}%
  \BibitemOpen
  \bibfield{author}{%
  \bibinfo {author} {\bibfnamefont{A.}~\bibnamefont{Saenz}},\ }%
  \bibfield{journal}{%
  \bibinfo {journal} {Phys.\,Rev.\ A}\ }%
  \textbf{\bibinfo {volume} {66}},\ \bibinfo {pages} {063407} (\bibinfo {year}
  {2002})%
  \bibAnnoteFile{NoStop}{sfm:saen02a}%
\bibitem{dia:sanc97b}%
  \BibitemOpen
  \bibfield{author}{%
  \bibinfo {author} {\bibfnamefont{I.}~\bibnamefont{S{\'a}nchez}}\ and\
  \bibinfo {author} {\bibfnamefont{F.}~\bibnamefont{Mart{\'\i}n}},\ }%
  \bibfield{journal}{%
  \bibinfo {journal} {J.\,Chem.\,Phys.}\ }%
  \textbf{\bibinfo {volume} {106}},\ \bibinfo {pages} {7720} (\bibinfo {year}
  {1997})%
  \bibAnnoteFile{NoStop}{dia:sanc97b}%
\bibitem{dia:sanc97a}%
  \BibitemOpen
  \bibfield{author}{%
  \bibinfo {author} {\bibfnamefont{I.}~\bibnamefont{S{\'a}nchez}}\ and\
  \bibinfo {author} {\bibfnamefont{F.}~\bibnamefont{Mart{\'\i}n}},\ }%
  \bibfield{journal}{%
  \bibinfo {journal} {J.\,Phys.\ B}\ }%
  \textbf{\bibinfo {volume} {30}},\ \bibinfo {pages} {679} (\bibinfo {year}
  {1997})%
  \bibAnnoteFile{NoStop}{dia:sanc97a}%
\bibitem{dia:bate51}%
  \BibitemOpen
  \bibfield{author}{%
  \bibinfo {author} {\bibfnamefont{D.~R.}\ \bibnamefont{Bates}},\ }%
  \bibfield{journal}{%
  \bibinfo {journal} {J.\,Chem.\,Phys.}\ }%
  \textbf{\bibinfo {volume} {19}},\ \bibinfo {pages} {1122} (\bibinfo {year}
  {1951})%
  \bibAnnoteFile{NoStop}{dia:bate51}%
\bibitem{sfm:lin00c}%
  \BibitemOpen
  \bibfield{author}{%
  \bibinfo {author} {\bibfnamefont{J.~T.}\ \bibnamefont{Lin}}\ and\ \bibinfo
  {author} {\bibfnamefont{T.~F.}\ \bibnamefont{Jiang}},\ }%
  \bibfield{journal}{%
  \bibinfo {journal} {Phys.\,Rev.\ A}\ }%
  \textbf{\bibinfo {volume} {63}},\ \bibinfo {pages} {013408} (\bibinfo {year}
  {2000})%
  \bibAnnoteFile{NoStop}{sfm:lin00c}%
\end{thebibliography}
%

%
%
\end{document}